\documentclass[twocolumn,pre,superscriptaddress,aps,10pt,floatfix]{revtex4-2}
\usepackage[utf8]{inputenc}
\usepackage[T1]{fontenc}
\usepackage[pdftex]{graphicx}
\usepackage[usenames,dvipsnames,svgnames]{xcolor}
\usepackage{amsmath}
\usepackage{amsfonts}
\usepackage{amssymb,comment}
\usepackage{physics}
\usepackage{braket}
\usepackage{dsfont}
\usepackage{bbold}
\usepackage[colorlinks,bookmarks=false,citecolor=NavyBlue,linkcolor=Red,urlcolor=blue]{hyperref}
\usepackage{tikz}
\usetikzlibrary{arrows.meta,calc,positioning}
\usepackage{orcidlink}
\usepackage{MnSymbol}
\usepackage[export]{adjustbox}
\newcommand*{\mc}{\mathcal}

\usepackage{thmtools}
\usepackage{thm-restate}

\newtheorem{thm}{Theorem}

\newcommand*{\nn}{\nonumber}
\newcommand{\Haar}{{\text{Haar}}}
\newcommand{\Wg}{{\text{Wg}}}
\newcommand{\Id}{\mathbb{1}}
\newcommand{\id}{\mathbb{1}}
\newcommand{\PauliX}{{\text{X}}}
\newcommand{\PauliY}{{\text{Y}}}
\newcommand{\PauliZ}{{\text{Z}}}
\newcommand{\Ex}{\mathbb{E}}

\newcommand{\numberofnoise}{n_{\mathcal{N}}} 

\newcommand{\new}[1]{\textcolor{black}{#1}}

\begin{document}

\title{{Noise-induced Simulability Transition from Operator Scrambling}}

\author{Neil Dowling~\orcidlink{0000-0002-4502-5960}}
\affiliation{Institut f\"ur Theoretische Physik, Universit\"at zu K\"oln, Z\"ulpicher Strasse 77, 50937 K\"oln, Germany}

\author{Xhek Turkeshi~\orcidlink{0000-0003-1093-3771}}
\affiliation{Institut f\"ur Theoretische Physik, Universit\"at zu K\"oln, Z\"ulpicher Strasse 77, 50937 K\"oln, Germany}
\affiliation{Institute of Quantum Control, Peter Grünberg Institut (PGI-8),
Forschungszentrum Jülich GmbH, 52425 Jülich, Germany}

\author{Jacopo De Nardis~\orcidlink{0000-0001-7877-0329}}
\affiliation{Laboratoire de Physique Th\'eorique et Mod\'elisation, CNRS UMR 8089, CY Cergy Paris Universit\'e, 95302 Cergy-Pontoise Cedex, France.}
\affiliation{JEIP, UAR 3573 CNRS, Collège de France, PSL Research University, 11 Place Marcelin Berthelot, 75321 Paris Cedex 05, France.}

\author{Guglielmo Lami~\orcidlink{0000-0002-1778-7263}}
\affiliation{Laboratoire de Physique Th\'eorique et Mod\'elisation, CNRS UMR 8089, CY Cergy Paris Universit\'e, 95302 Cergy-Pontoise Cedex, France.}
\affiliation{JEIP, UAR 3573 CNRS, Collège de France, PSL Research University, 11 Place Marcelin Berthelot, 75321 Paris Cedex 05, France.}

\begin{abstract}
The complexity of simulating quantum many-body dynamics, or quantum computations, in the Heisenberg picture is governed by the scrambling of initially simple operators into superpositions of exponentially many Pauli strings. \new{The coefficients of the corresponding expansion define the Pauli spectrum, whose structure controls the performance of classical algorithms based on truncating in the Pauli basis.} Here we determine the finite-depth Pauli spectrum of random quantum circuits, both in the noiseless case and in the presence of local depolarizing noise, through its moments, given by the operator stabilizer R\'enyi entropies. In noiseless circuits, we uncover a hierarchy in the approach to the fully scrambled regime: low moments equilibrate at relatively short depths, while higher moments, which are sensitive to rare, large-amplitude Pauli coefficients, require parametrically larger depths. In noisy circuits, scrambling competes with an effective suppression of operator spreading. Above a critical error per cycle $\gamma_c N =\mathcal{O}(1)$, the operator fails to reach the fully scrambled distribution and remains supported on an atypically sparse subset of Pauli strings. Conversely, below this scale, we rigorously show that simulation via Pauli propagation techniques remains exponentially hard, demonstrating that {finite noise does not automatically imply classical simulability}. \new{Our results point to a universal noise-induced transition in operator complexity, separating intrinsically quantum operator dynamics from classically tractable regimes.} 
\end{abstract}

\maketitle
\section{Introduction}
Quantum processors can generate and time-evolve quantum many-body states whose dynamics rapidly escapes efficient classical description. Local information spreads across an exponentially large Hilbert space, making direct classical simulation intractable. In the Schr\"odinger picture, this complexity is reflected for example in the rapid growth of entanglement~\cite{Calabrese_2005,Nahum2017}; in the Heisenberg picture, it appears instead as the \emph{scrambling} of an initially simple observable into a superposition of exponentially many Pauli operators~\cite{Larkin1969QuasiclassicalMI,Shenker_Stanford_2014,Swingle2016,Nahum2018,Keyserlingk2018,Fisher2023}. Beyond its relevance for classical simulation, this operator proliferation is a ubiquitous signature of quantum chaos~\cite{Prosen2007,Roberts2016,Dubail_2017,Alba2019,dowling2023scrambling} and thermalization~\cite{Srednicki,Deutsch1991} in many-body systems.

Quantum processors are, however, inevitably {noisy}~\cite{Arute_2019,boixo2018characterizing,Morvan2023,Zhang2023,Quantinuum2025,Liu2025,Yin2025,Haghshenas2025,Shirizly2024,Smith2025,Rost2025,Ringbauer2025,Abanin2025,Manetsch2025}. Depolarization, dephasing, and amplitude damping degrade quantum information and suppress operator scrambling. The competition between coherent scrambling and noise-induced damping is central to both fundamental many-body physics and the near-term prospects for quantum advantage. If noise suppresses scrambling strongly enough, dynamics that would be intractable in the absence of noise may become efficiently approximable by classical algorithms~\cite{Aharonov2023NoisyRandomCircuitSampling}. Several recent developments have clarified aspects of this phenomenon: algorithms based on truncating the expansion of Heisenberg-evolved operators in the basis of Pauli operators have proved remarkably effective~\cite{Rakovszky2022,Pollmann2022,begusic2024realtime,angrisani2024classically,rudolph2025pauliprop}, especially in noisy systems~\cite{schuster2024polynomialtime,Angrisani2025ArbitraryLocalNoisePauliPropagation,j1gg-s6zb}. In parallel, studies of open quantum circuits have shown that noise suppresses the contribution of early circuit layers, so that local expectation values and output statistics depend only on an effective shallow depth~\cite{mele2024noisei,sauliere2025noisyq,wei2026noise}, while related work has identified the emergence of approximate Markovian structure~\cite{zhang2025classicallysamplingnoisyquantum}. 

\begin{figure}[t!]
    \centering
    \includegraphics[width=0.95\linewidth]{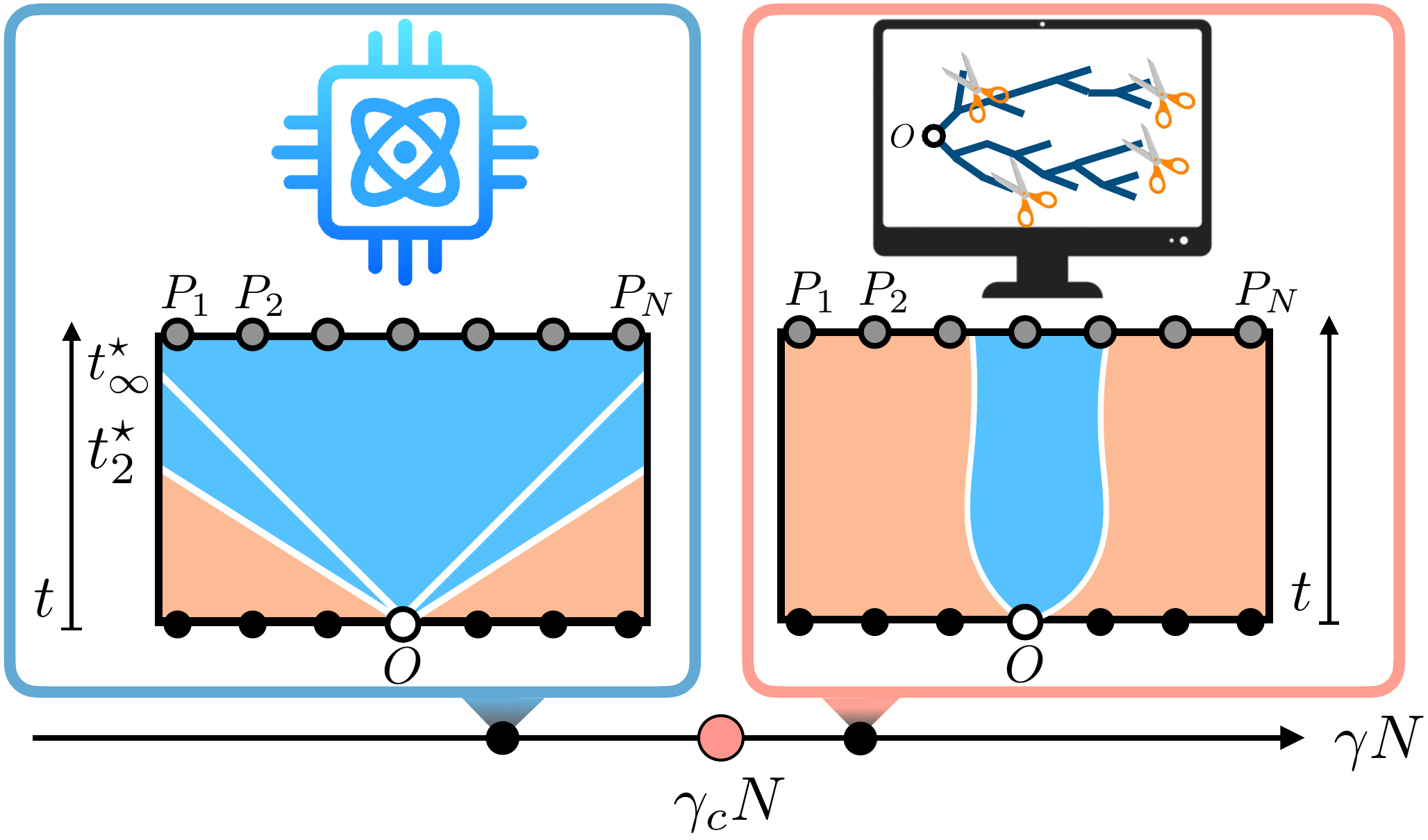}
    \caption{Schematic evolution of the moments of the Pauli spectrum, equivalently the operator stabilizer R\'enyi entropies. In the unitary case (left), different moments equilibrate at distinct crossover times $t_2^\star < t_3^\star < \ldots < t_\infty^\star$, revealing a hierarchy in the approach to the fully scrambled, Haar-random regime. In the noisy case (right), increasing the error per cycle $\gamma N$ arrests this flow and drives the operator into a sparse, classically simulable phase that can be captured by Pauli truncation algorithms.}
    \label{fig:cartoon}
\end{figure}

What is still lacking is a microscopic understanding of how noise and unitary evolution jointly shape the internal structure of an operator, together with a quantitative criterion for \textit{whether and when} a critical noise strength exists beyond which this structure becomes simple enough to be captured by classical algorithms based on Pauli-path truncation.

In this work, we address both questions \textit{via three complementary approaches}: we solve a simpler random tensor network ansatz, we analyze a statistical mechanics mapping of brickwork random circuits, and perform extensive numerical simulations.

First, we obtain a complete characterization of the Pauli spectrum of an initially local operator scrambled by random quantum circuits, a paradigmatic setting that captures universal features of chaotic many-body dynamics. The moments of this spectrum reveal a sharp \emph{hierarchy of scrambling times}, with higher moments requiring greater circuit depth to equilibrate. These time scales differ from the conventional design time of random circuits: whereas full operator scrambling requires depths extensive in the number of qubits, approximate designs are reached at only logarithmic depth~\cite{Harrow2009,Brand_o_2016,Chen_2025,laracuente2026shallowapproximateunitarydesigns,schuster2025randomunitariesextremelylow}.
Second, we study the competition between noise and unitary scrambling. Specifically, \emph{we identify a critical error rate per cycle below which scrambling dominates}, driving the operator toward a fully scrambled regime in which the Pauli spectrum approaches a Porter--Thomas-like distribution (see Fig.~\ref{fig:cartoon}). Above this threshold, noise instead arrests the spreading of operator weight through Pauli space, leaving behind a sparse backbone of dominant Pauli strings (composed mostly of identities), signaled by heavy tails in the Pauli spectrum. \new{We further argue that this structural change in the Pauli spectrum coincides with a \emph{sharp transition in classical simulability}---at least according to sparse Pauli representation-based methods. Specifically, we rigorously show that below the transition, no classical simulation scheme based on truncating the time-evolved operator in the Pauli basis can efficiently achieve a small relative error. Thus, the presence of finite noise does not, by itself, imply classical simulability, suggesting that the effect of weak noise on classical simulability may be more subtle than indicated by some recent results. Above the transition, in contrast,  Pauli-truncation-based simulations become efficient.  We provide numerical evidence of this transition in terms of the relative error of a standard Pauli propagation algorithm for random/typical initial states. } 
 
\new{To arrive at our results, we derive exact analytical results for a simplified model of noisy random circuits, the random matrix product unitary (RMPU), which captures the universal features of operator scrambling. We identify a heuristic mapping from RMPUs to conventional local circuits, arguing that such random tensor networks are well-representative of such random circuit models. Indeed, we show that the RMPU reproduces the same behavior found through an independent statistical mechanics argument for the conventional brickwork random circuit model. Supporting this picture, the analytical results are in excellent agreement with extensive numerical simulations.}

\section{Pauli spectrum and simulability} \label{sec:pp}
Any operator $O$ acting on a system of $N$ qubits, with total Hilbert space dimension $D = 2^N$, admits a unique decomposition in the Pauli basis
\begin{equation}
    O=\sum_P a_P P
    \qquad  a_P=\frac{\Tr(OP)}{D} \, , \label{eq:pauli_expan}
\end{equation}
where $P\in\{\Id,\PauliX,\PauliY,\PauliZ\}^{\otimes N}$ are strings of Pauli operators. By squaring the coefficients $a_P$, one can define a probability distribution over the Pauli strings:
\begin{equation}\label{eq:pi}
    \pi_O(P)
    =
    \frac{a_P^2}{\| O\|_2^2 }
    =
    \frac{1}{D} \frac{\Tr[O P]^2}{\Tr[O^2]} \, ,
\end{equation}
where the extra factors ensure the correct normalization $\sum_P \pi_O(P) = 1$, and $\| O\|_2:= \sqrt{\Tr[O^2]/D}$ is the normalized Hilbert-Schmidt norm of $O$. The distribution $\pi_O(P)$ carries detailed information on the structure of the operator. Specifically, to quantify the spread of $O$ in the Pauli basis, one can use the $k^{\mathrm{th} }$ R\'enyi entropy
\begin{equation}\label{eq:OSE}
    M^{(k)}(O)
    =
    \frac{1}{1-k}
    \log\left(\sum_P \pi_O(P)^k\right) \, ,
\end{equation}
also known as \emph{operator stabilizer R\'enyi entropies} (OSE)~\cite{dowling2024magicheis}. The full set of OSEs, across all R\'enyi indices $k$, can in turn be encoded in a single object: the \emph{Pauli spectrum}. Defined as
\begin{equation}\label{eq:Pi_u}
\Pi_O(u) = D^{-2}\sum_P \delta\!\left(u-D^2\pi_O(P)\right) \, ,
\end{equation}
it is a probability distribution in $u$ that assigns a delta peak to all (rescaled) values of $\pi_O(P)$. Its $k^{\mathrm{th} }$ moment is proportional to the argument of the log in the $k^{\mathrm{th} }$ OSE:
\begin{equation}
\mu_k(O)=\int_0^{\infty} du \, u^k \Pi_O(u)=D^{2k-2}\sum_P \pi_O(P)^k \, . \label{eq:moments}
\end{equation}
For a simple local operator $O = Z_j$ (with $j \in {1, \dots, N}$), the Pauli spectrum is $\Pi_{O}(u) = (1 - D^{-2}) \, \delta(u) + D^{-2} \,\delta(u - D^{2})$, with exponentially large moments $\mu_k(O) = D^{2k-2}$. Under typical local quantum dynamics, however, the operator spreads, the spectrum broadens, and the moments decay. At large depth, the operator becomes effectively fully Haar scrambled. In this limit, the Pauli moments are of order one, specifically ${\mu}_k(O) \to \overline{\mu_k}^{\mathrm{Haar}} := (2k-1)!!$~\cite{dowling2024magicheis}, corresponding to the distribution $\Pi_{\mathrm{OPT}}(u) = {e^{-u/2}}/({\sqrt{2\pi u}})$. We refer to this as the \emph{Operator Porter--Thomas} (OPT) distribution, by analogy with the standard Porter--Thomas law for output probabilities of quantum states in random circuits~\cite{Porter1956,TurkeshipauliPT2025}. 

Informally, order-one values of the Pauli moments $\mu_k$ indicate that exponentially many Pauli operators contribute comparably, while large values indicate the presence of a persistent backbone of dominant Pauli operators.
The latter regime is precisely the one exploited by \emph{Pauli-propagation algorithms}~\cite{Rakovszky2022,Aharonov2023NoisyRandomCircuitSampling,begusic2024realtime,schuster2024polynomialtime,angrisani2024classically,Shao2024NoisyVQAPolynomial,rudolph2025pauliprop,
GonzalezGarcia2025PauliPathBeyondAverage,
Fontana2025NoisyVQALOWESA,
Rudolph2023LOWESASurrogate,
Angrisani2025ArbitraryLocalNoisePauliPropagation} which approximate time-evolved operators by keeping only $N_P$ terms in the decomposition of Eq.~\eqref{eq:pauli_expan}. 

The Pauli moments directly bound the possibility of accurate and efficient simulations via Pauli propagation~\cite{dowling2024magicheis,dowling2026classicalsim}. \new{Specifically, if one truncates $O$ to $\tilde O = \sum_{j=1}^{N_P} a_{P_j} P_j$, then there exists a state $\rho$ for which the error in the expectation value is lower bounded by the $k$-th OSE for any $k\geq1$, as~\cite{dowling2024magicheis}
\begin{equation}\label{eq:finalTrunc}
    |\tr[(O-\tilde O)\rho]|
    \geq
    \frac{\| O\|_2 }{2N}
    \left(
    M^{(k)}(O)-\log (N_P)-1
    \right)  .
\end{equation}
Thus, if the OSE scales as $M^{(2)}(O)=\Omega(N)$, then achieving a target approximation error $\epsilon$ requires $N_P$ to grow super-polynomially with $N$; equivalently, expectation values of $O$ cannot be efficiently simulated by Pauli truncation. 
Conversely, in sub-extensive regimes, one expects efficient simulability, although the above bound alone does not rigorously establish it. This could in principle be shown -- at least for correlations on-average over one-designs or high-temperature states -- by considering Rényi entropies with $k<1$~\cite{Schuch_2008,dowling2026classicalsim}, which are, however, inaccessible via the replica methods we adopt in the following sections.}
\new{The above establishes that understanding the behavior of the OSE, or equivalently of the Pauli moments $\mu_{k}(O)$, as a function of circuit depth and error per cycle is essential for quantitatively identifying the boundary between classically simulable and intractable dynamics based on sparse Pauli representations. We refine Eq.~\eqref{eq:finalTrunc} in the presence of noise later in Theorem~\ref{thm:sim}.}

\new{\section{Universal scrambling times in local circuits}}
In this Section we will study the behavior of the Pauli spectrum from random quantum circuits without noise, revealing a operator space anticoncentration phenomena. \\

\subsection{A solvable model of operator scrambling}

\begin{figure}[t!]
    \centering
    \includegraphics[width=0.95\linewidth]{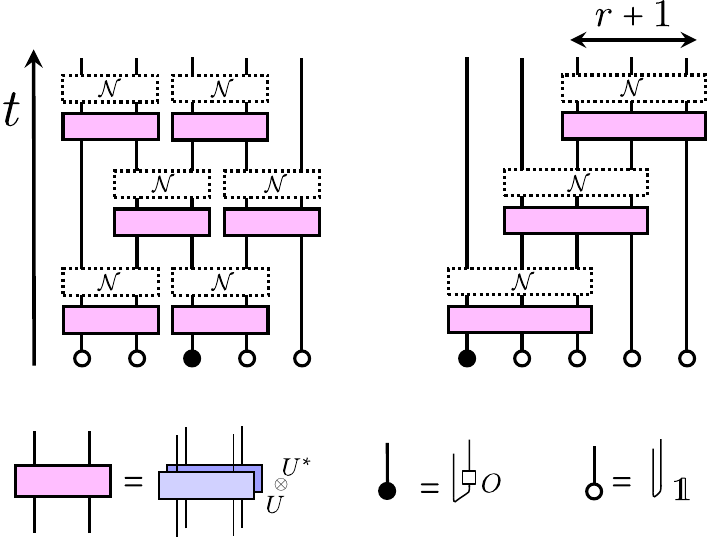}
    \caption{\textbf{Brickwork random circuits and RMPU.} (Left) In a brickwork circuit, an initially local operator spreads through successive two-site gates $U$ as the circuit depth $t$ increases. (Right). A random matrix product unitary is constructed by overlapping Haar-random blocks, each acting on a support of size $r+1$, with neighboring blocks sharing an overlap of dimension $r$. Both geometries can possibly be affected by noise, modeled by a noise channel $\mathcal{N}$ acting after each unitary gate (dotted boxes).}
    \label{fig:architectures}
\end{figure}

We first solve an analytically tractable model of scrambling in which an initially localized operator $O$ evolves under a \emph{Random Matrix Product Unitary} (RMPU) $U$ (see Fig.~\ref{fig:architectures}, right). An RMPU corresponds to a staircase circuit of Haar-random gates, each acting on $r+1$ qubits, with consecutive gates overlapping on $r$ sites. The parameter $r$ controls how much the operator scrambles per circuit layer. Equivalently, the final operator $O_U = U O U^{\dag}$ can be represented as a Matrix Product Operator (MPO) with bond dimension $\chi^2$, where $\chi=d^r$ and we often choose qubits, $d=2$. This construction preserves the essential finite-depth structure of local chaotic circuits (see Fig.~\ref{fig:architectures}, left) while allowing certain exact computation~\cite{Lami2025}.

Using techniques from the Weingarten calculus to analytically perform the average over the Haar-random gates~\cite{collins_integration_2006,Kostenberger_2021,Collins_2022}, the averaged Pauli moments $\overline{\mu_k}:=\Ex_U[\mu_k (O_U) ]$ can be rewritten as a product of transfer matrices in a space of size $(2k)!$, multiplied by vectors at the boundaries (see \new{Appendix~\ref{ap:rmpu} for details}). The exact evaluation of the dominant terms of this product allows us to obtain (for sufficiently large $N$ and $\chi$)
\begin{equation}\label{eq:unitaryRMPU}
    \overline{\mu_k}
    =
     \overline{\mu_k}^{\rm Haar}
    \left[
    1+
    C_k
    \left(
    \frac{d^{N(1-k^{-1})}}{\chi}
    \right)^{2k}
    \right] \, ,
\end{equation}
with $C_k=({d^2-1})/({d^{2k}-d^2})$. The overall prefactor is the Haar value (corresponding to the OPT distribution), while the second term is the leading finite-depth correction. The latter depends on system size and bond dimension only through the scaling variable $d^{N(1-1/k)} / \chi$. This fact immediately implies a hierarchy of scrambling thresholds: the second moment reaches its Haar value when $\chi\gg d^{N/2}$; the third requires $\chi\gg d^{2N/3}$; and the $k^{\mathrm{th}}$ moment equilibrates only when $\chi\gg d^{N(1-1/k)}$.

\begin{figure}[t]
    \centering
    \includegraphics[width=0.95\linewidth]{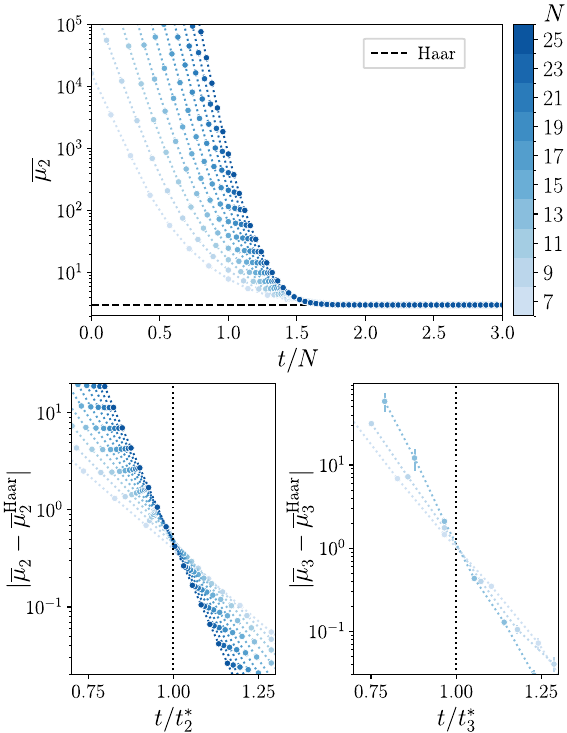}
    \caption{\textbf{Equilibration of Pauli moments in noiseless 1D brickwork circuits}. (Top) Second moment $\overline{\mu_2}$ of the Pauli spectrum as a function of the circuit depth $t$. The black line is the Haar value $\overline{\mu_2}^{\rm Haar}$. (Bottom) Deviation of the second ($\overline{\mu_2}$, left) and third ($\overline{\mu_3}$, right) moments from their respective Haar values $\overline{\mu_2}^{\mathrm{Haar}}$ and $\overline{\mu_3}^{\mathrm{Haar}}$. On the $x$-axis, the circuit depth is rescaled by the timescales of Eq.~\eqref{eq:timescales} with $\tau$ given by Eq.~\eqref{eq:tau}. Different sizes $N$ define a crossover occurring at depth $t/t_k^\star\simeq1$, consistent with Eq.~\eqref{eq:unitaryRUC}. \new{Data are obtained using the replica tensor-networks for the second moment, and exact diagonalization for the third moment ($3 \cdot 10^3$ circuit realizations); see Appendix~\ref{ap:numerics} for more details}.}
    \label{fig:plots1}
\end{figure}

\begin{figure*}[t!]
    \centering
    \includegraphics[width=1.01\linewidth]{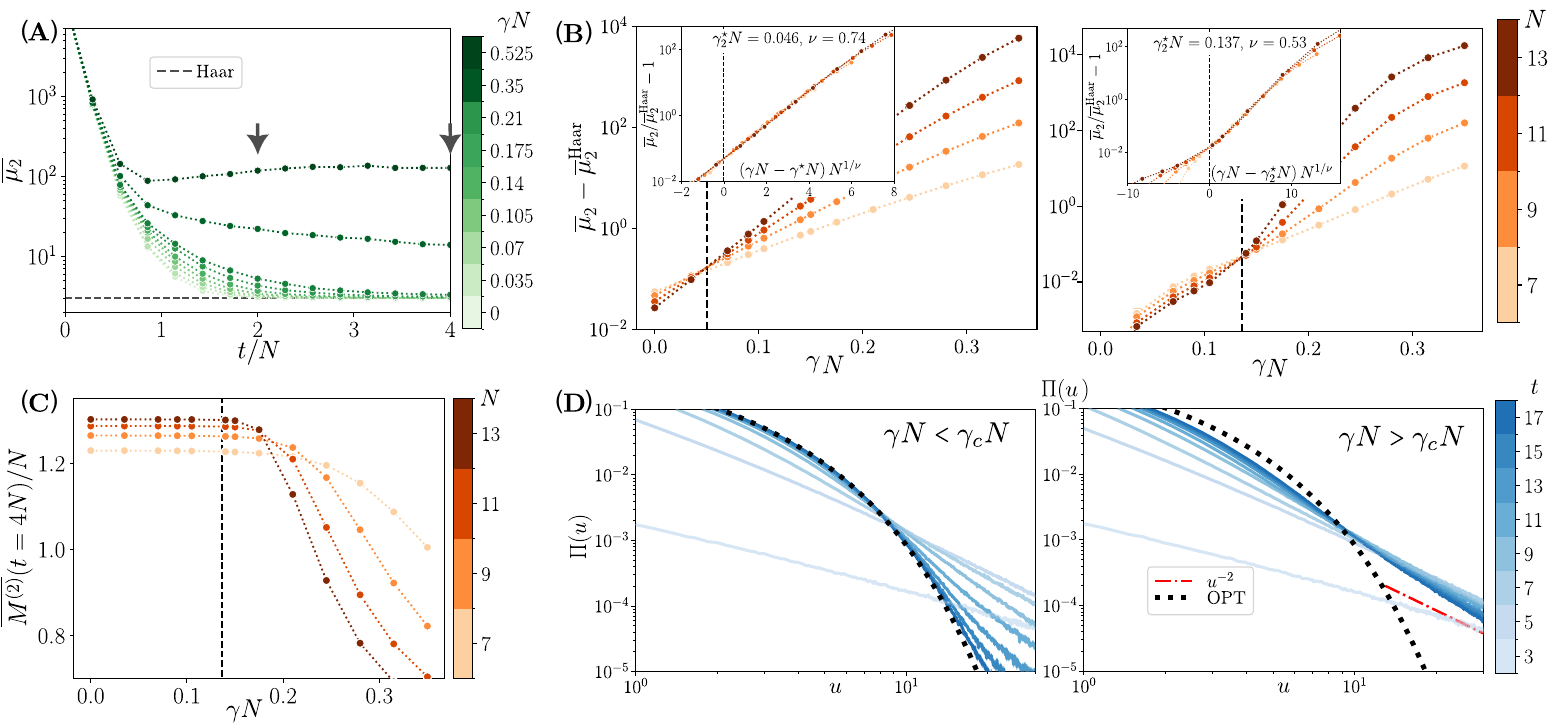}
    \caption{\textbf{Pauli spectrum and its moments in noisy circuits.} (A) Average normalized second moment $\overline{\mu}_2$ for noisy Haar-random brickwork circuits with $N=7$ qubits and varying error per cycle $\gamma N$. For sufficiently weak noise, $\mu_2$ asymptotically approaches the Haar value (dashed line), whereas above a noise threshold it saturates at a value larger than the Haar prediction. Arrows indicate the values of depth considered for the next sub-figures. (B) Deviation from Haar of the second moment, $\overline{\mu}_2-\overline{\mu}_2^{\Haar}$, at depth $t=2N$ (panel on the left) and $t=4N$ (panel on the right) vs $\gamma N$, for different $N$. Below the predicted onset $\gamma_2^\star N$ (vertical dashed line, Eq.~\eqref{eq:shifted_transition_point}), the deviation goes to zero as $N$ increases; above it, the deviation grows exponentially with $N$. Inset: collapse of $\overline{\mu}_2/\overline{\mu}_2^{\rm Haar}-1$ vs $(\gamma N-\gamma_2^\star N)\,N^{1/\nu}$. The fit gives $\gamma_2^\star N\simeq0.046$ for $t=2N$ and $\gamma_2^\star N\simeq0.137$ for $t=4N$ (with effective $\nu\simeq0.74$ and $0.53$), in good agreement with the prediction of Eq.~\eqref{eq:shifted_transition_point}, $\gamma_2^\star N\simeq0.050$ and $0.137$, respectively. 
    (C) Accordingly, the average 2-OSE $\overline{M^{(2)}}/N$ at depth $t=4N$ shows the same onset. Below $\gamma_2^\star N$ (vertical dashed line), the OSE is maximal, $\overline{M^{(2)}}/N\simeq 2\ln 2$. Above it, the OSE density decreases with $\gamma N$, and the OSE becomes subextensive at $\gamma_c N=\tau^{-1}$.(D) Pauli spectrum $\Pi_O(u)$ of an operator initially localized at the central site of a $3\times3$ lattice, evolved by a 2D noisy brickwork circuit of depth $t$. Below the transition ($\gamma N\simeq0.28$, left), the distribution converges to the OPT distribution. Above it ($\gamma N\simeq1.05$, right), it stays far from the OPT, with persistent heavy tails $\sim u^{-2}$. The 1D case (not shown) behaves analogously.}
    \label{fig:plots2}
\end{figure*}
\new{\subsection{Mapping to the brickwork circuit}}
We now connect our exact RMPU results to conventional local circuits and to their paradigmatic random counterparts, i.e.\ random brickwork circuits. We first note that time evolution under a local circuit turns initially local operators into MPOs whose bond dimension $\chi(t)$ grows exponentially with the depth $t$~\cite{Vidal2004TEBD}, whereas the RMPU construction yields MPOs with bond dimension $\chi^2$. This motivates the identification (see also Ref.~\cite{sauliere2025noisyq} and Fig.~\ref{fig:architectures})
\begin{equation}\label{eq:chitot}
    \chi \propto  e^{t/\tau} \, ,
\end{equation}
where $\tau$ is a microscopic, model-dependent parameter, interpreted as a scrambling time. The variable $\tau$ can, in principle, depend on the moment index $k$; however, our numerics (see below) suggest that it is in fact independent of $k$ and relates to the entangling power of the circuit~\cite{sauliere2025noisyq}. Substituting Eq.~\eqref{eq:chitot} into Eq.~\eqref{eq:unitaryRMPU} gives the universal scaling form
\begin{equation}\label{eq:unitaryRUC}
    \overline{\mu_k}
    \simeq
     \overline{\mu_k}^{\rm Haar}
    \left[
    1+
    C_k'
    \left(
    \frac{d^{N(1-k^{-1})}}{e^{t/\tau}}
    \right)^{2k}
    \right] \, ,
\end{equation}
where $C_k'$ is now an unknown, non-universal constant. Eq.~\eqref{eq:unitaryRUC} implies a hierarchy of operator scrambling times in terms of circuit depth. Specifically, define
\begin{equation}\label{eq:timescales}
    t_k^\star
    =
    N   \tau (1-k^{-1})\log d  \, .
\end{equation}
For $t<t_k^\star$, the finite-depth correction grows exponentially with system size; for $t>t_k^\star$, it is exponentially suppressed. This defines a crossover that sharpens as the system size $N$ increases. In this sense, $t_k^\star$ is the depth at which the time-evolved operator converges to the OPT distribution up to the $k$-th moment of the Pauli spectrum. The hierarchy of these scrambling times implies that while the bulk of the Pauli spectrum converges to the OPT distribution already at low depths ($t \sim t_2^*$), the \emph{heavy tails}, responsible for the higher moments, persist and converge only at longer times. 

In Fig.~\ref{fig:plots1}, we test this prediction in Haar-random brickwork circuits by numerically evaluating the moments $\mu_k$ using tensor-network contractions in replica space and direct Pauli sampling. We find that the timescale $\tau$ agrees with the one governing the decay of the half-system purity in random unitary circuits. The latter can be computed analytically in $(1+1)$-dimensional circuits using $k=2$ replicas, yielding~\cite{Nahum2017}
\begin{equation}\label{eq:tau}
  \tau^{-1} = \log\left((d^2+1)/(2d)\right) .
\end{equation}
Substituting this value and using Eq.~\eqref{eq:timescales}, we find a sharp crossover at $t/t_k^\star \simeq 1$. The agreement indicates that the RMPU via the identification Eq.~\eqref{eq:chitot} captures the universal finite-depth approach to the Haar-scrambled regime.

\new{\section{Noise-induced Pauli spectrum transition}}

We next consider the presence of local depolarizing noise and its effect on operator scrambling, a noise model that is representative of generic features of weak unital noise~\cite{Dalzell2024}. The depolarizing channel $\mathcal{N}_{\gamma}(O) = (1-\gamma) O + \Id  \cdot \Tr[O]/ \Tr[\Id]$ rescales each non-identity Pauli operator by $(1-\gamma)$, leaving the identity unchanged. Thus, while unitary dynamics scrambles the operator over Pauli space, noise locally suppresses non-identity components. 
Notice also that, unlike in the unitary case where $O_U^2=\Id$ and $\Tr[O_U^2]=D$, noise makes the normalization in Eq.~\eqref{eq:pi} depend non-trivially on $U$. We therefore define unnormalized moments ${\nu}_k(O) := D^{-2} \sum_P \Tr[OP]^{2k}$. The normalized moments can then simply be recovered as $\mu_k(O)={\nu}_k(O) / \big({\nu}_1(O) \big)^k$, \new{with ${\nu}_1(O) = \|O\|_2^2$};
the averaged unnormalized moments are $\overline{\nu_k}:=\Ex_U[{\nu}_k(O_U)]$. \new{In the following theorem, we refine Eq.~\eqref{eq:finalTrunc} to the case of noisy circuits; specifically, we make the bound depend explicitly on the averaged unnormalized moments $\overline{\nu_k}$, which we are in turn able to compute analytically.}

\new{
\begin{thm} \label{thm:sim}
    Let $U$ be a random quantum circuit of arbitrary geometry on $N$ qudits of local dimension $d$, containing $\numberofnoise$ depolarizing channels $\mathcal N_\gamma$. Let $O$ be an initially normalized operator, $\|O\|_2=1$, with $O_U$ its Heisenberg-evolved form, and let $\tilde{O}_U$ be an approximation to $O_U$ consisting of $N_P$ Pauli strings. Then, for every integer $k\ge 2$, the circuit-averaged normalized error is lower bounded in terms of the averaged unnormalized Pauli moments $\overline{\nu_k}$ as 
    \begin{align}\label{eq:finalTruncThm}
        \begin{split}
        & \Ex_U \bigg[\frac{ {\sup_{\rho}|\tr[(O_U-\tilde O_U)\rho]|}}{\|O_U\|_2}\bigg] \geq \\
        & \quad \frac{1}{2N}
        \left(
        \frac{1}{1-k}\log \left( \frac{\overline{\nu_k}}{F^{2k} d^{N(2k-2)}} \right) - \log N_P -1
        \right)  ,            
        \end{split}
    \end{align}
    for $k\ge 2$. Here, $F\equiv(1-\gamma)^{\numberofnoise}\le\|O_U\|_2$ is the overall circuit fidelity, and the supremum is over all initial states for each sample. 
\end{thm}
This result follows closely from Ref.~\cite{dowling2024magicheis}, and is detailed in Appendix~\ref{ap:sim}. Notice that $\numberofnoise=m=N-r$ for the noisy RMPU model, while $\numberofnoise = Nt$ in the brickwork circuit after $t$ layers.} 

\new{To interpret the above Theorem, we recall the strong notion of simulability in terms of Pauli propagation~\cite{dowling2026classicalsim} as described in Section~\ref{sec:pp}: when an approximation $\tilde{O}_U$ retains only $N_P=\mathrm{poly}(N)$ Pauli strings while reproducing expectation values over \textit{all} initial states up to inverse polynomial error, then we say that $O_U$ is efficiently approximated via Pauli propagation. Theorem~\ref{thm:sim} then means that for sufficiently weak noise---where the moments are close to the OPT distribution and so the right-hand side of Eq.~\eqref{eq:finalTruncThm} is $\mathcal{O}(1)$---each sampled circuit is not simulable according to Pauli propagation, on average across circuit realizations. The focus of this Section is to make this statement precise; we will uncover how weak, exactly, the noise channels need to be to produce this non-simulability guarantee, and when there is a transition to an `easy' regime. We will return to the question of simulability in more detail in Section~\ref{sec:simtransition}.}

\new{We further point out that the left-hand side of Eq.~\eqref{eq:finalTruncThm} is a normalized error: the factor
$\|O_U\|_2$ accounts for the decrease of the operator's overall norm due to the noise channels, i.e.\ both numerator and denominator typically decrease exponentially with the number of depolarizing channels. This normalization does not account for the fact that, after anticoncentration due to the unitary dynamics, typical expectation values are exponentially smaller than $\|O_U\|_2$, so that resolving them requires an exponentially small absolute error. Later in Section~\ref{sec:simtransition} we instead numerically examine the error relative to the typical size of expectation values, and find that the noise-induced simulability transition persists according to relative error, and is thus meaningful also beyond the anticoncentration depth.} 

{\subsection{Noisy RMPU: exact solution and mapping to brickwork circuits}}

\new{While we will return to Theorem~\ref{thm:sim} and simulability according to Pauli propagation in Section~\ref{sec:simtransition}, we first proceed to evaluate the averaged unnormalized Pauli moments $\overline{\nu_k}$ for the RMPU model. Even with noise, the RMPU model is still fully analytically solvable.}

Including a depolarizing channel after every unitary gate (see Fig.~\ref{fig:architectures}, right), we find
\begin{align}\label{eq:formula_noise_neilnew}
    \overline{\nu_k}
    &=
    F_{\rm RMPU}^{2k}
     \, \overline{\mu_k}^{\rm Haar}
    \left[
    1+
    C_k(\gamma)
    \left(
    \frac{d^{N(1-k^{-1})}}
    {\chi \, F_{\rm RMPU}}
    \right)^{2k}
    \right] \, ,
\end{align}
where $F_{\rm RMPU}=(1-\gamma)^m$ is the overall circuit fidelity, $m=N-r$ the total number of gates, and $C_k(\gamma) = \left( d^2-1 \right) / \left( d^{2k} (1-\gamma)^{-2k} - d^2 \right)$. The prefactor arises from the decay of the unnormalized operator norm due to noise. \\

To connect this \new{exact} result \new{for RMPU} with the brickwork \new{geometry}, we again employ the mapping in Eq.~\eqref{eq:chitot}. Moreover, we also replace the RMPU fidelity with the analogous circuit fidelity in the brickwork case, 
\begin{equation}
    F = (1-\gamma)^{Nt} \simeq e^{-(\gamma N) t},
\end{equation}
with error per cycle $\gamma N$. \new{We therefore arrive at}
\begin{equation}\label{eq:scaling_noisy}
    \frac{\overline{\nu_k}}{F^{2k}}
    \simeq
    \overline{\mu_k}^{\rm Haar}
    \left[
    1
    +
    C_k(\gamma)
    \left(
    \frac{
    e^{\gamma N t}
    d^{N(1-k^{-1})}}
    {e^{t/\tau}}
    \right)^{2k}
    \right],
\end{equation}
with \new{(likely non-universal) positive constants $C_k(\gamma)=\mathcal{O}(1)$. Eq.~\eqref{eq:scaling_noisy} can alternatively be derived heuristically by directly mapping the brickwork circuit to a statistical-mechanical model; see Section~\ref{sec:stat_mech}.}
The correction to Haar contains two competing terms: unitary scrambling contributes $e^{-t/\tau}$ and suppresses the correction, while noise contributes $e^{\gamma N t}$ and amplifies it. Their balance defines the critical error per cycle
\begin{equation}\label{eq:threshold}
    \gamma_c N = \frac{1}{\tau} \, . 
\end{equation}
For $\gamma N < \gamma_c N$, the correction decays with depth on a time scale renormalized by the noise
\begin{equation}
    t^{\gamma, \star}_k = t_k^\star / (1 -  \gamma N \tau) \, ,
\end{equation}
which diverges for $\gamma \to \gamma_c$. \new{Beyond $t^{\gamma,\star}_k$, the $k$-th moment approaches its Haar value exponentially fast in depth. Since $t^{\gamma,\star}_k$ increases with $k$, the full Pauli spectrum follows the OPT law for all $t \gtrsim t^{\gamma,\star}_\infty = N\tau\ln d/(1-\gamma N\tau)$.}
\new{For $\gamma >\gamma_c $, the correction instead grows exponentially with depth, and consequently the Pauli spectrum remains distinct from the OPT at all depths. In this case, the correction to Haar term of Eq.~\eqref{eq:scaling_noisy} has the form of the contribution of only a few ($\mathcal O(1)$) Pauli strings that together carry a fraction $\exp(-2t/\tau(1-\gamma N \tau ))$ of the operator weight. Above the threshold, this fraction grows with depth, because noise suppresses operators faster than scrambling can build them up. This growth is ultimately limited by $\mu_k(O) \le d^{N(2k-2)}$, the extreme case of an operator made of a single Pauli string. In practice, the Pauli spectrum converges to a stationary distribution different from the OPT law: a heavy-tailed one, in which a few strings carry a finite fraction of the weight.} \\

\subsection{Statistical-mechanics interpretation}\label{sec:stat_mech}
\new{In this Section, we support the scaling form of the Pauli moments, Eq.~\eqref{eq:scaling_noisy}, derived from the RMPU solution, with a simple statistical-mechanics interpretation of the brickwork circuit~\cite{Nahum2017,Nahum2018,Keyserlingk2018,PhysRevB.99.174205,PhysRevX.10.031066,ChanDeLucaChalker2018PRL,PhysRevX.8.031057}}. Indeed, applying a standard approach for random circuits, averaging over the Haar-random gates in the brickwall circuit to compute the $k^{\mathrm{th} }$ Pauli moment results in an Ising-like model. This model features permutation variables $\sigma \in S_{2k}$ interacting via three-body interactions on a triangular lattice, as:
\begin{equation}
    \mathbb{E}_U \Bigg[ \Big(\includegraphics[scale=0.30, valign=c]{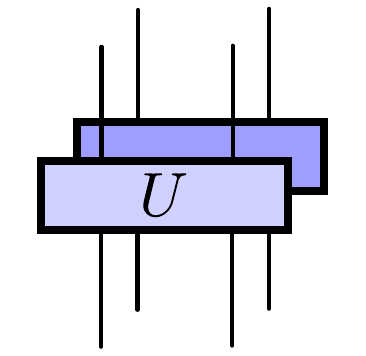} \Big)^{\otimes 2k} \Bigg] \, \, \, \longrightarrow \, \, 
    \begin{tikzpicture}[scale=1.2, baseline=15pt]
\coordinate (A) at (-0.5,0);
\coordinate (B) at (0.5,0);
\coordinate (C) at (0,+0.866); 
\draw[thick, fill=green!20] (A) -- (B) -- (C) -- cycle;
\filldraw[fill=white, draw=black] (A) circle (1.5pt);
\filldraw[fill=white, draw=black] (B) circle (1.5pt);
\filldraw[fill=white, draw=black] (C) circle (1.5pt);
\node[scale=0.8] at (-0.5,-0.15) {$\sigma_1$};
\node[scale=0.8] at (0.5,-0.15) {$\sigma_2$};
\node[scale=0.8] at (0,+1.0) {$\sigma_3$};
\node[scale=0.8] at (0.03,+1.3) {$J(\sigma_1,\sigma_2,\sigma_3)$};
\end{tikzpicture}
\, ,
\end{equation}
\new{where the interaction strength $J(\sigma_1,\sigma_2,\sigma_3)$ can be derived from the Weingarten calculus under unitary Haar averaging (see Appendix~\ref{ap:stat_mech} for details).}
The moments $\mu_k$ correspond to the partition function of this model (strictly speaking, this holds only for sufficiently large local dimension $d$~\cite{Bao_2020}). Inspection of the weights $J(\sigma_1,\sigma_2,\sigma_3)$ shows that the model is ferromagnetic, with domain walls incurring a suppression factor of $\sim d^{-1}$ for each unit cell they pass through~\cite{PhysRevB.99.174205}. Moreover, information causality gives rise to a lightcone, outside of which the permutation variables are necessarily the identity; we describe this in detail in Appendix~\ref{ap:stat_mech}. In the absence of noise ($\gamma=0$) and at sufficiently large depth, the dominant contribution to the partition function comes from ferromagnetic uniform configurations within this lightcone, where the permutations correspond to perfect matchings of the $2k$ replicas (i.e.\ ways of grouping them into $k$ pairs). These configurations are schematically shown in Fig.~\ref{fig:cartoon} (left), and their degeneracy,  corresponding to the total number of pairings $(2k-1)!!$, yields the Haar value $\overline{\mu}_k^{\mathrm{Haar}}$.

Non-vanishing noise ($\gamma > 0$) acts as a sort of external magnetic field that biases the statistical weights toward the identity permutation~\cite{PhysRevB.108.104310,sauliere2025noisyq}. Consequently, the configurations considered above are suppressed by a factor $e^{-\gamma N t \alpha_k}$, where $\alpha_k N t$ comes from the area of the region with non-identity permutations, with constant $\alpha_k=\mc{O}(1)$ . These configurations now compete with other terms coming from pairings that remain confined within a region of size $\mc{O}(1)$ throughout the entire evolution, as depicted in the right panel of Fig.~\ref{fig:cartoon}. These other configurations are suppressed by a (relative) factor of $e^{-t/\tilde{\tau}_k} \, e^{N \beta_k}$ due to vertical domain walls of length $\sim t$ (compared to the previous case where the domain walls had length $\sim \beta_k N$). Both $\tilde{\tau}_k$ and $\beta_k$ are constants. Combining all these contributions, and including a prefactor $C_k$, we find
\begin{equation}\label{eq:stat_mec}
    \overline{\nu_k} \simeq \overline{\mu_k}^{\mathrm{Haar}} e^{- \gamma N t \alpha_k} \left[ 1 + C_k \, e^{+\gamma N t \alpha_k} e^{-t/\tilde{\tau}_k} e^{N \beta_k}  \right].
\end{equation}
This form is identical to that found in Eq.~\eqref{eq:scaling_noisy} upon identifying $\beta_k=2k(1-k^{-1})\log d$, $\alpha_k=2k$ and $\tilde{\tau}_k= \tau/(2k)$. Therefore, the statistical-mechanical picture of the brickwork circuit fully confirms our scaling prediction.
\new{This picture also clarifies the nature of the transition. Indeed, at fixed depth $t \propto N$, one has, up to $N$-independent prefactors, $\overline{\nu_k}$ is the sum of two terms correspond to distinct domain-wall configurations. Each term depends smoothly on $\gamma N$. The transition occurs when they exchange dominance, so it is first order. Correspondingly,
the free energy $-N^{-1}\ln\overline{\nu_k}$ has a kink at
\begin{equation}\label{eq:shifted_transition_point}
    \gamma_k^{\star} N = \frac{1}{\tilde{\tau}_k \alpha_k} - \frac{\beta_k}{c\, \alpha_k}
    = \frac{1}{\tau} \left(1 - \frac{t^{\star}_k}{t} \right),
\end{equation}
which tends to $\gamma_c N = 1/\tau$ for $t \to \infty$. At finite $N$, this kink is smoothed over a range of $\gamma N$ shrinking as $1/N$.}

\new{\subsection{Numerical evidence}}

We test the predictions above \new{for the Pauli spectrum numerically}, using Haar-random brickwork circuits with error strength $\gamma$ per gate (see Fig.~\ref{fig:plots2}). \new{We evolve the operator exactly in the Heisenberg picture, which gives direct access to the normalized Pauli moments $\mu_k$, and average over circuit realizations.}
First, we observe that at weak noise, the second moment $\overline{\mu_2}$ converges to the Haar value, as in the noiseless case (Fig.~\ref{fig:plots2} (A)). At stronger noise, it instead saturates above the Haar value, indicating that the operator never reaches the fully scrambled regime. 
\new{At fixed depth, $t=2N$ or $t=4N$, the deviation from Haar,
$\overline{\mu}_2-\overline{\mu}_2^{\Haar}$, decreases with $N$ for weak noise and grows with $N$ for
stronger noise, for $N$ from $7$ to $13$ (see Fig.~\ref{fig:plots2}(B)). The curves for different $N$ cross at a noise strength in good agreement with $\gamma_2^\star N$ from  Eq.~\eqref{eq:shifted_transition_point} ($0.050$ and $0.137$, respectively).} \new{Similarly, at depth $t=4N$, the $2$-R\'enyi OSE density stays maximal up to the onset predicted by Eq.~\eqref{eq:shifted_transition_point}, and decreases beyond it, with a crossover that sharpens with $N$ (Fig.~\ref{fig:plots2} (C)).} \new{Remarkably, the observed onset agrees with Eq.~\eqref{eq:shifted_transition_point}, with $\tau$ given by Eq.~\eqref{eq:tau}.} The asymptotic threshold, $\gamma_c N=\tau^{-1}$, coincides with the noise threshold found in cross-entropy benchmarking (XEB) studies of noisy random circuits~\cite{ware2023sharpphasetransitionlinear,Morvan2023,sauliere2025noisyq}. This suggests that the XEB transition and the loss of Pauli-based simulability are two manifestations of the same
phenomenon: a noise-induced reduction of the effective depth. \\
In \new{Fig.~\ref{fig:plots2} (D), we also extend} our numerics to 2D, showing the Pauli spectrum $\Pi_O(u)$ of a time-evolved operator in a noisy random brickwork circuit. There, we also observe the predicted behavior: for small error per cycle, convergence to the OPT; for large error values, persistent heavy tails. \\

\begin{figure*}[t!]
    \centering
    \includegraphics[width=1.0\linewidth]{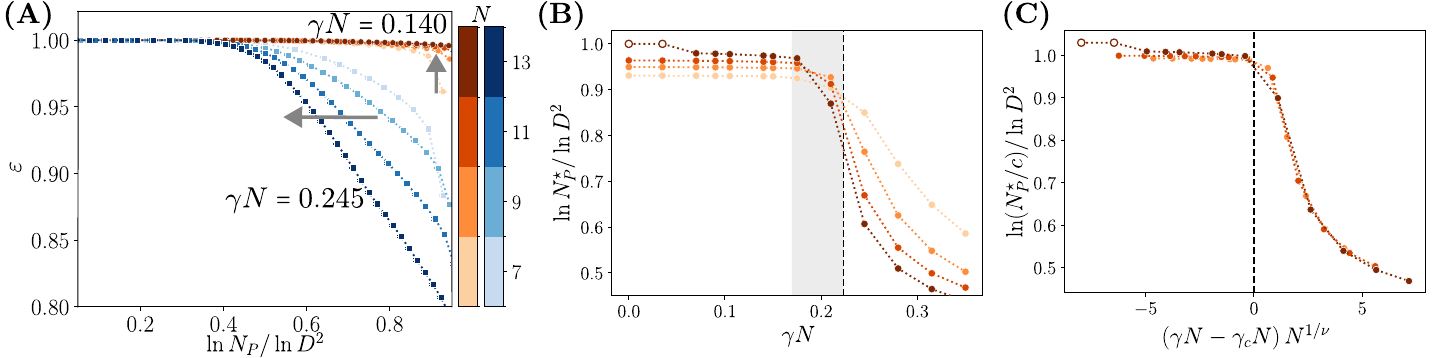}
    \caption{\textbf{Pauli-truncation simulability transition.} \new{Evolution of a local operator from noisy Haar random brickwork circuits of depth $t=4N$ for $N=7$--$13$ with Pauli propagation, keeping the $N_P$ largest Pauli components after each gate. We consider the Hilbert--Schmidt error averaged over circuit realizations $\varepsilon=\Ex_U[\varepsilon_U]$ with $\varepsilon_U$ defined in Eq.\eqref{eq:ratio_new}, and approximately $ 3.6 \cdot 10^3$ realizations of the circuit $U$. (A)~Error $\varepsilon$ versus $\ln N_P/\ln D^2$, below (orange, $\gamma N=0.14$) and above (blue, $\gamma N=0.245$) the critical noise. With increasing $N$, below the threshold the error approaches one for any fixed exponent, while above the threshold the curves move towards a vanishing value of $N_P/D^2$ (see grey arrows). (B)~We define $N_P^\star$ as the smallest Pauli budget achieving $\varepsilon\le 1-\delta$ ($\delta=0.05$ in the plot). Open markers indicate that the threshold is not reached within the largest value of $N_P$. Below the threshold, $\ln N_P^\star/\ln D^2$ remains close to one, i.e.\ $N_P^\star\propto D^2$; above it, it decreases with $N$. The shaded region is the expected finite-size shift, $\ln\delta^{-1}/(2t)$, obtained by requiring the weight of the few dominant Pauli strings, $\sim \exp\big(-2t/\tau(1-\gamma N \tau ) \big)$, to exceed $\delta$. (C)~Finite-size collapse of $\ln(N_P^\star/c)/\ln D^2$, where $c$ is the prefactor of $N_P^\star\simeq c\,D^2$ on the hard plateau, for $N=9-13$, with fitted $\gamma_c N\simeq 0.184$ and $\nu\simeq0.68$.}}
    \label{fig:plots3}
\end{figure*}

\section{Pauli-truncation simulability transition}\label{sec:simtransition}
\new{We now return to the question of \emph{simulability} by algorithms based on Pauli truncation~\cite{Aharonov2023NoisyRandomCircuitSampling,schuster2024polynomialtime,Shao2024NoisyVQAPolynomial,GonzalezGarcia2025PauliPathBeyondAverage,j1gg-s6zb,Angrisani2025ArbitraryLocalNoisePauliPropagation,Nelson2026NaturallyFaultTolerant}. Pauli propagation is generally based on iteratively evolving then discarding strings in the operator's Pauli decomposition, Eq.~\eqref{eq:pauli_expan}, based on number of identity components or on the size of the coefficients, $|a_P|$. \\
First, we shall remark that our rigorous bound, Eq.~\eqref{eq:finalTruncThm}, combined with the Pauli-spectrum transition found above, points towards a simulability transition. We notice that the right-hand side of Eq.~\eqref{eq:finalTruncThm} equals $\log d-\log(\overline{\nu}_k/F^{2k})/[2N(k-1)]-\log N_P/(2N)$ (up to $\mathcal O(1/N)$ terms). Inserting the random circuit moments expression, Eq.~\eqref{eq:scaling_noisy}, the logarithm is set by the larger of its two terms: the Porter--Thomas term and the correction due to the few dominant Pauli strings. Therefore, we obtain
\begin{equation}
    \begin{split}
        \label{eq:theoremreformulated}
    & \Ex_U \bigg[ \frac{{\sup_{\rho}|\tr[(O_U-\tilde O_U)\rho]|}}{\|O_U \|_2}\bigg]  \\
    & \quad \gtrsim  \min\!\big[\log d,\;\tfrac{k}{k-1}\,\tfrac{t}{\tau N}\,(1-  \gamma N \tau )\big]-\ln N_P/(2N).
    \end{split}
\end{equation}
For $\gamma<\gamma_k^\star$, the minimum on the right-hand side equals $\ln d$; for $\gamma_k^\star<\gamma<\gamma_c$, it equals $\tfrac{k}{k-1}\,\tfrac{t}{\tau N}\,(1-\gamma N\tau)$, which
is positive and of order one at any depth $t\propto N$; for $\gamma>\gamma_c$, it becomes negative. Hence, for $\gamma<\gamma_c$, the worst-case error of expectation values (normalized by $F$), remains of order one for any approximation of $O_U$ by $N_P=\mathrm{poly}(N)$ Pauli strings. For $\gamma<\gamma_k^\star$, the same holds even when $N_P$ is exponentially large in $N$, as long as it is an exponentially small fraction of all $D^2$ Pauli strings (since the bound holds for every $k\ge 2$, the widest range is obtained for $k=2$, for which
$\gamma_k^\star$ is largest). Consequently, below the threshold, no algorithm based on a Pauli-sparse representation of the evolved operator, such as Pauli propagation with any truncation rule, can be
efficient in the worst case. }

\new{Theorem~\ref{thm:sim} establishes therefore hardness below the threshold. For $\gamma>\gamma_c$, however, the right-hand side of Eq.~\eqref{eq:finalTruncThm} can become negative already for polynomial $N_P$, so the bound gives no
information. We therefore test Pauli propagation numerically on noisy brickwork circuits, fixing the number $N_P$ of retained Pauli strings and comparing with the exact evolution ($N_P=D^2$); see Fig.~\ref{fig:plots3}. To quantify the error, we average the relative squared error of expectation values over Haar-random pure states $\rho_\psi=|\psi\rangle\langle\psi|$. For traceless operators, this average reduces to the relative Hilbert--Schmidt distance between the operator $O_U$ and its truncation $\tilde O_U$, i.e.
\begin{align}
  \mathbb E_{\psi \sim \mathrm{Haar}} \Big[ \frac{\mathrm{tr}[(\tilde O_U - O_U) \rho_\psi]^2}
     {\mathrm{tr} [O_U\rho_\psi]^2} \big] &\simeq \frac{\mathbb E_{\psi}\big[\mathrm{tr}[(\tilde O_U - O_U) \rho_\psi]^2\big]}
     {\mathbb E_{\psi }\big[\tr[O_U\rho_\psi]^2\big]} \nonumber \\
&=  \frac{\|\tilde O_U-O_U\|_2^2}{\|O_U\|_2^2} =:\varepsilon_U,\label{eq:ratio_new}
\end{align}
which we will further average over circuit realizations $U$. The reason we average numerator and denominator separately in Eq.~\eqref{eq:ratio_new}, is that the ratio $\mathbb E_\psi\big[\langle\psi|\tilde O_U-O_U|\psi\rangle^2/\langle\psi|O_U|\psi\rangle^2\big]$ can generally diverge. Indeed, since $O_U$ is traceless, the distribution of $\langle\psi|O_U|\psi\rangle$ has a nonzero density at zero, while the numerator generically does not vanish for the same states. This remains true even when $\tilde O_U$ keeps the $N_P$ largest Pauli coefficients of $O_U$. The ratio of averages in Eq.~\eqref{eq:ratio_new} is nevertheless the typical value of the ratio: for large $D$, $\langle\psi|\tilde O_U-O_U|\psi\rangle$ and $\langle\psi|O_U|\psi\rangle$ are jointly Gaussian with zero mean, and for such variables the median of the ratio of their squares equals exactly the ratio of their second moments (see Appendix~\ref{sec:haaraveragerelativeerror} for further details). }

\new{We numerically study the mean square relative error in Fig.~\ref{fig:plots3}, averaging Eq.~\eqref{eq:ratio_new} over the circuit instances $U$, $\varepsilon := \Ex_U[\varepsilon_U]$. In Fig.~\ref{fig:plots3} (left) we plot $\varepsilon$ as a function of
$\ln N_P/\ln D^2$ for two values of $\gamma N$, below and above the transition, where the curves for different $N$ behave differently. For $\gamma N<\gamma_c N$, the curves approach $1$ as $N$
increases, while for $\gamma N>\gamma_c N$ they shift towards smaller $\ln N_P/\ln D^2$. Hence, above the transition, an exponentially small fraction of all Pauli strings suffices to capture a finite fraction of the operator's influence. This is further witnessed by the center and right panels, where we define $N_P^\star$ as the smallest $N_P$ achieving $\varepsilon\le 1-\delta$ ($\delta=0.05$ in the figure). As a function of $\gamma N$,
$\ln N_P^\star/\ln D^2$ shows a crossover that sharpens with $N$, consistent with the predicted transition at $\gamma_c N=\tau^{-1}$.} 

\new{Finally, we stress that our notion of non-simulability is worst-case: Theorem~\ref{thm:sim} (and consequently Eq.~\eqref{eq:theoremreformulated}) guarantees a large error for some state $\rho$, which may be chosen adversarially and depend on the circuit instance (not necessarily for a fixed reference state). Our results therefore complement, and are consistent with, previous work on simulability guarantees for Pauli propagation algorithms. For instance, Ref.~\cite{angrisani2024classically} proves efficient additive-error classical approximability of expectation values in (zero noise) random circuits. However, this is according to the weaker notion of a fixed reference state and does not involve any normalization by circuit fidelity, in contrast to Eq.~\eqref{eq:finalTruncThm}.}

\section{Discussion}
\new{The proposed noise-induced simulability transition} connects with several recent results which showed efficient classical sampling and simulability of expectation values in the cases of constant error strength $\gamma$~\cite{schuster2024polynomialtime,mele2024noisei,zhang2025classicallysamplingnoisyquantum}. Our results instead identify a lower noise scale: \new{we have shown that in random circuits, below a critical noise level ($\gamma < \gamma_c = \mathcal{O}(N^{-1})$), expectation values of time-evolved operators are generically \emph{not} efficiently simulable according to Pauli propagation algorithms,} i.e.\ they come with an exponential cost. We therefore establish that \emph{finite noise does not automatically guarantee classical simulability}. Given that some current quantum platforms are well below this critical value~\cite{Morvan2023,Abanin2025,Manetsch2025,HangleiterQuantumAdvantage}, this observation is particularly relevant for sampling noisy circuits. Several results have shown that in the presence of a finite Markov length, classical sampling is possible~\cite{zhang2025markovianity,zhang2024nonlocal,wei2026mie,napp2022efficient,lee2025correlation}. However, this requires simulating a large 2D network under weak noise up to a relative precision, which may still be computationally hard according to the noise transition.  

Our results also have implications beyond simulability. First, the Pauli spectrum refines conventional probes of quantum chaos, including out-of-time-ordered correlators~\cite{Nahum2018,Xu2022-ue} and operator entanglement entropies~\cite{Prosen2007,Dubail_2017,Alba2019,Dowlin2024LOE-OSRE}, by resolving not only the growth of operator support but also the distribution of weights within that support. 
Moreover, moments of the Pauli spectrum also quantify the operator's non-stabilizerness~\cite{dowling2024magicheis}, i.e.\ its ``magic'' resource, a necessary ingredient for quantum advantage distinct from entanglement~\cite{gottesman1998,Aaronson2004} and of recent interest in the context of many-body physics~\cite{Liu_2022,Leone2022stab,turkeshi2024magicspreadingrandomquantum}. Their finite-depth hierarchy offers a fine-grained measure of the build-up of non-stabilizer resources in noisy and near-term devices, and so may offer insight into (measurement-induced) phase transitions~\cite{White2021Conformal,Niroula2024}. A natural next step is to study the Pauli spectrum in magic-constrained settings, e.g.\ circuits dominated by Clifford gates, such as error-correction codes~\cite{PhysRevLett.102.110502}, or in more complex noise models, such as under non-unital~\cite{PRXQuantum.5.030317} or non-Markovian errors~\cite{White2025nonMark}. Our analytic methods are likely amenable to the non-unital case using the Pauli representation thereof~\cite{mele2024noisei}. Clarifying whether an analogous noise-induced transition persists in such dynamics would sharpen the link between operator scrambling, magic generation, and classical simulability.

\newpage
\appendix 
\onecolumngrid 
\section*{Appendix}
\twocolumngrid

\section{Pauli Propagation Simulability from  Operator Stabilizer Entropies} \label{ap:sim}
\new{We first give some extra details on the relation between the moments of the Pauli spectrum and simulability according to Pauli truncation schemes. This provides extra details on the background of Section~\ref{sec:pp} as well as a proof of Theorem~\ref{thm:sim}.}

From a result in Ref.~\cite{dowling2024magicheis}, we know that for any Hermitian operator $O$ with norm $\|O \|_2=1$, 
\begin{equation} \label{eq:sim1}
    \| O - \tilde{O} \|_\infty \geq \frac{1 }{2N}
    \left(
    M^{(1)}(O)-\log (N_P)-1
    \right) \,  ,
\end{equation}
where $\tilde{O}$ is constructed from the Pauli strings with the $N_P$ largest coefficients in the expansion Eq.~\eqref{eq:pauli_expan}, and $\|X\|_\infty $ is the spectral norm, i.e.\ the largest singular value of $X$. To arrive at Eq.~\eqref{eq:finalTrunc}, we first replace $M^{(1)}(O)$ with $M^{(2)}(O)$, through the inequality between R\'enyi entropies: $M^{(1)}(O) \geq M^{(k)}(O)$ for $k\geq 1$. Next, we note that $O - \tilde{O} = \sum_{j = N_P +1}^{D^2} a_{P_j} P_j$ is Hermitian (since Pauli strings are Hermitian and the $a_{P_j}$ are real). Therefore, the largest singular value of $O - \tilde{O}$ is real and corresponds to the eigenvector $\ket{\lambda_{\mathrm{max}}}$. By choosing $\rho = \ket{\lambda_{\mathrm{max}}}\! \bra{\lambda_{\mathrm{max}}}$, we obtain: $| \tr[(O - \tilde{O})\rho] | = \| O - \tilde{O} \|_\infty $. Finally, considering an unnormalized $O$, we recall a key step in the proof of Eq.~\eqref{eq:sim1}: $\| O - \tilde{O} \|_\infty \geq \sqrt{\sum_{j = N_P + 1}^{D^2} a_{P_j}^2}$. The lower bound in Eq.~\eqref{eq:sim1} then follows from the fact that $\sum_{j = N_P + 1}^{D^2} a_{P_j}^2$ is the tail sum of a classical distribution. Therefore, after normalization, the proof proceeds as usual for normalized coefficients, but with an overall factor of $\sqrt{\sum_{j = 1}^{D^2} a_{P_j}^2} = \| O \|_2$, leading to the final result in Eq.~\eqref{eq:finalTrunc}. We also note that retaining the largest $N_P$ Pauli terms is the optimal truncation, and other Pauli truncation schemes (e.g., based on the size of the non-identity support of the strings) may only worsen the approximation~\cite{rudolph2025pauliprop}. \\

To apply Eq.~\eqref{eq:finalTrunc} to our results on the average-case Pauli spectrum of random circuits, we rewrite it under an averaging as
\new{
\begin{align}\label{eq:finalTrunc1}
\begin{split}
     & \Ex_U \bigg[ \frac{\sup_\rho|\tr[(O_U-\tilde O_U)\rho]|}{\|O_U\|_2} \bigg]  \\
    &\!\geq \! 
    \frac{1}{2N} \!
    \left( \frac{1}{1-k}
    \overline{\log(D^{-2k+2} \mu_k) }-\log (N_P)-1\!
    \right)\! \\
    & \geq
    \frac{1}{2N}
    \left(
    \frac{1}{1-k}\log(D^{-2k+2} \overline{\mu_k}) -\log (N_P)-1
    \right)
\end{split}
\end{align}
}
where we have applied Jensen's inequality for the negative logarithm. 
The left-hand side is the relative error in expectation values, and $\overline{\mu_k}$ is the quantity we compute for ensembles of noiseless random circuits; for noisy circuits, instead, we compute the averaged, unnormalized moments $\overline{\nu_k}$. To relate the latter to the bound above, we notice that $\mu_k(O) = \nu_k(O) / \big(\nu_1(O)\big)^k$, with $\nu_1(O)=\Tr[O^2] / D = \| O\|_2^2$, and therefore: 
\begin{align}
    \begin{split}
    -\log \overline{\mu_k} \geq -\log \left( \frac{\overline{\nu_k}}{(\underset{U}{\inf} \|O_U\|_2)^{2k} } \right)\geq -\log \left( \frac{\overline{\nu_k}}{(1-\gamma)^{2 \numberofnoise k}} \right) \, ,         
    \end{split}
\end{align}
where we have first used the positivity of $\nu_k$, and then specified 
\new{to a circuit with a total of $\numberofnoise$} depolarizing channels, each of which may reduce the Hilbert-Schmidt norm of the input by at most $(1-\gamma)$. 
\new{Notice that, $\numberofnoise=m=N-r$ for the noisy RMPU model, while $\numberofnoise = Nt$ in the brickwork circuit after $t$ layers.}
\new{We are therefore assured that, when the quantity
$\overline{\nu_k}/(1-\gamma)^{2\numberofnoise k}$ remain of order one, or more generally grow slower
than their maximal value $d^{2N(k-1)}$, the operator cannot be
approximated by $\mathrm{poly}(N)$ Pauli strings according to any (possibly adversarially chosen) initial state, and so Pauli
propagation is not efficient. Putting the above together, we arrive at Theorem~\ref{thm:sim}.}

\section{Proof of Replica Tensor Network Results} \label{ap:rmpu}
Here we supply proofs of the analytic results of the main text on the Pauli spectrum of the random tensor network model, with and without depolarizing noise channels.

\subsection{Replica approach}
Consider an initial Pauli operator $O$ acting on a single site of a $D = d^N$-dimensional Hilbert space $\mathcal{H}$. We examine its Heisenberg evolution $O_U = U O U^{\dag}$ under a random unitary circuit $U$, which we first take to be noiseless. Our quantity of interest is the average $k^{\mathrm{th}}$ moment of its Pauli spectrum (Eq.~\eqref{eq:moments}),
\begin{equation} \label{eq:mu1}
    \overline{\mu_k} := \Ex_U[{\mu}_k(O_U)]= D^{-2} \sum_{P \in \mc{P}_N} \Ex_U[ \Tr[UOU^{\dag} P]^{2k} ] \, , 
\end{equation}
where $\Ex_U[\cdot]$ denotes averaging over a random circuit ensemble (e.g.\ brickwork or RMPU, see Fig.~\ref{fig:architectures}), and the sum runs over all Pauli strings $\mc{P}_N = \{\Id, X, Y, Z\}^{\otimes N}$. In the following, we require the technical assumption that the local dimension is a power of $2$ (i.e., multiple qubits), $d=2^p$, such that the Pauli operators are Hermitian. To analytically compute these moments, we adopt a replica approach. First, via the Choi isomorphism, any operator $O$ can be identified with an (unnormalized) quantum state, i.e.\ $|O \rrangle := (O \otimes \Id) \sum_i \ket{ii}$, where the sum runs over a basis $\{ \ket{i} \}$ of $\mathcal{H}$. Slightly abusing notation, we write $O \equiv O \otimes \id^{N-1}$ for the local operator embedded in the full Hilbert space. Vectorizing both $O$ and the Pauli operators $P$, Eq.~\eqref{eq:mu1} becomes
\begin{equation}\label{eq:mukreplica}
 \overline{\mu_k} = D^{-2} \sum_{P \in \mc{P}_N}  \llangle P |^{\otimes 2k} \, \Ex_U[(U \otimes U^*)^{\otimes 2k}] \, | O \rrangle^{\otimes 2k} \, .
\end{equation}
Here, we have isolated the average over $2k$ copies of $U$ and $U^*$, so that we can first compute it, then contract the resultant expression with the left and right boundary conditions of $\sum_{P}  \llangle P |^{\otimes 2k} $ and $| O \rrangle^{\otimes 2k}$, respectively. Note, also, that the Pauli strings in the sum factorize over qudits: 
\begin{equation}\label{eq:pauli_factorize}
    \sum_{P \in \mc{P}_N}  \llangle P |^{\otimes 2k}  = \bigotimes_{i=1}^N \left( \sum_{P_i \in \mc{P}_1 }  \llangle P_i |^{\otimes 2k} \right) \, . 
\end{equation}
As the random circuits we study consist of gates that are uniformly sampled over the unitary group, we now employ the Weingarten calculus.

\subsection{Weingarten calculus}
This mathematical toolbox expresses the Haar average $\Ex_{U\sim \Haar(q)}[\dots]$ of $n$ copies of the unitary gates $U, U^\dagger$, each acting on a Hilbert space of dimension $q$, as a sum over permutation operators between the copies of Hilbert space. For a $q$-dimensional space with basis states $\{ |i \rangle \}_{i=0}^{q-1}$, the permutation operators $\sigma \in S_n$ act as: $\sigma \ket{i_1 \dots i_n} = \ket{i_{\sigma(1)} \dots i_{\sigma(n)}}$. Then, using the vectorized representation described above, the Weingarten formula reads~\cite{collins_integration_2006,Collins_2022}
\begin{equation}\label{eq:weingarten}
\mathbb{E}_{U \sim \Haar(q)} \left[ (U \otimes U^*)^{\otimes n} \right] = \sum_{\pi, \sigma \in S_n} \Wg_{\pi  \sigma}(q)\, |\pi\rrangle \llangle \sigma| \, .
\end{equation}
Here, $\Wg(q)$ denotes the $n! \times n!$ Weingarten matrix, defined as the (pseudo)inverse of the matrix of overlaps between permutations $G_{\pi,\sigma}(q):=\llangle \pi |\sigma\rrangle = q^{\#(\pi^{-1}\sigma)}$. Here, $\#(\cdots)$ counts the number of cycles in a permutation, e.g., $\sigma = (12)(3) \in S_3$ corresponds to a SWAP between the first two replicas and identity on the third, and $\#(\sigma)=2$. Note also from the Choi isomorphism that the inner product between vectorized operators is given by $\mathrm{tr}(O^\dagger Q) = \llangle O | Q \rrangle$. The Weingarten matrix $\Wg(q)$ is generally difficult to compute for arbitrary replicas $n$. Since we need information for all $n$ to access the full Pauli spectrum, we use its asymptotic expansion. To leading order in $1/q$,~\cite{Collins_2022}
\begin{equation}\label{eq:Wgexp}
\mathrm{Wg}_{\pi,\sigma}(q) = \frac{M(\pi^{-1}\sigma)}{q^{2n-\#(\pi^{-1}\sigma)}} \left(1 + \mc{O}(q^{-2}) \right) \,,
\end{equation}
where $M(\cdot)$ is a constant dependent on the cycle structure of its argument; relevant to us, $M(e) =1$ and $M(\tau) =(-1)^{n/2}$, where $e=(1)(2)(3)\dots (n)$ is the identity permutation and $\tau=(12)(34)\dots$ is a perfect pairing ($n/2$ transpositions). In the following, we will apply these formulas to the case of $n=2k$ replicas to evaluate Eq.~\eqref{eq:mukreplica}.

\subsection{Noiseless RMPU}
We now consider the random matrix product unitary (RMPU) circuit ensemble in the noiseless case ($\gamma=0$). RMPU are obtained as in Fig.~\ref{fig:architectures} (right) through a staircase circuit acting on a system of $N$, $d-$dimensional qudits (calculations presented in this and the next section are valid for any $d = 2^p$, for integer $p\geq 1$). Each gate acts on a space of dimension $d \chi = d^{r+1}$ (with $\chi=d^r$) and the total number of layers is $m=N-r$. The calculation of $\overline{\mu_k}$ boils down to using Eq.~\eqref{eq:weingarten} multiple times to perform the independent average over each unitary gate within Eq.~\eqref{eq:mukreplica}. The resulting expression can be recast as a one-dimensional product of matrices, projected onto vectors at the boundaries, namely
\begin{align}\label{eq:transfer_matrices}
\overline{\mu_k} &= \mathbb{E}_{U_i \sim \Haar(d \chi)} \Bigg[ \sum_{P_i}\Bigg(\includegraphics[scale=0.65, valign=c]{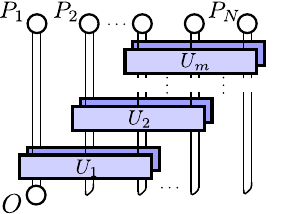} \Bigg)^{\otimes 2k} \Bigg] \nn \\
&=:(L)^{\mathsf{T}} T^{m-1} R \,.
\end{align}
The transfer matrix $T$ and boundary vectors $L$ and $R$ are defined in a $(2k)!$-dimensional space corresponding to $S_{2k}$. The boundary vector $L$ comes from contracting a permutation $\sigma$ with the initial operator: $L_{\sigma}= \llangle O^{\otimes 2k}| \sigma \rrangle = \prod_{c\in \text{cycles}(\sigma)} \Tr[O^{|c|}] = \chi^{\#(\sigma)} \mathbf{1}_{E}(\sigma)$, where $c$ runs over cycles of $\sigma$ (with length $|c|$), $\mathbf{1}_E(\sigma)$ is the indicator function for permutations with only even-length cycles, and we used $\Tr[O]=0$ and $O^2=\Id$. 
Similarly, permutations contracted with the Pauli operators $P$ at the end of the circuit give a weight: $d^{-2} \sum_P \llangle P^{\otimes 2k}| \sigma \rrangle = d^{\#(\sigma) + 2 \cdot \mathbf{1}_E(\sigma) - 2}$. In this way, we arrive at: $T=\Lambda_1(d)\,\mathrm{Wg}(d\chi)\,\Lambda_2(d)\,G(\chi)$, where we introduce the diagonal matrices $[\Lambda_1]_{\sigma\pi} = \delta_{\sigma \pi} d^{\#(\sigma)}$ (corresponding to the identity boundary) and $[\Lambda_2]_{\sigma \pi} = \delta_{\sigma \pi} d^{\#(\sigma) + 2 \cdot \mathbf{1}_E(\sigma) - 2}$ (corresponding to a single-qudit Pauli sum from Eq.~\eqref{eq:pauli_factorize}). The final boundary vector is $R_{\sigma} = \sum_{\pi \in S_{2k}}[\Lambda_1(d)]_{\sigma \sigma} [\Wg(d \chi)]_{\sigma \pi} [\Lambda_2(d)^{r+1}]_{\pi}$. 

We analyze Eq.~\eqref{eq:transfer_matrices} using a controlled expansion for $\chi \to \infty$. Since off-diagonal elements of $G(\chi)$ and $\Wg(d\chi)$ are suppressed by at least $\chi^{-1}$, we have
\begin{equation}\label{eq:transfer}
T_{\sigma \pi} = d^{a(\sigma)} \left( \delta_{\sigma \pi} + \mc{O}(\chi^{-1}) \right)
\end{equation}
with 
\begin{equation}
a(\sigma)=2\#(\sigma) + 2 \, \mathbf{1}_E(\sigma) - 2 - 2k \, .
\end{equation}
The dominant term in Eq.~\eqref{eq:transfer_matrices} therefore comes from taking the diagonal part of $T$, yielding: $\overline{\mu_k} = \sum_{\sigma \in S_{2k}} \mathbf{1}_{E}(\sigma) D^{a(\sigma)} + \mc{O}(\chi^{-1})$. Among the subset of even permutations, the weight $a(\sigma)$ is maximized by the pairings, i.e.\ permutations $\tau$ consisting of $k$ transpositions pairing distinct elements, for which $a(\tau)=0$. Consequently, for large $N$, this yields the Haar value: $\overline{\mu_k} = (2k-1)!! + \mc{O}(\chi^{-1}) + \mc{O}(D^{-1})$. 

To analyze the corrections to this asymptotic value, we first notice that among all permutations in $S_{2k}$, the weight $a(\sigma)$ is instead maximized by the identity permutation $e$. In fact: $a(e)=2k - 2$, which is positive for all $k \geq 2$. While the vector $L$ initializes the left boundary into even permutations, leading corrections therefore arise from using off-diagonal terms of the transfer matrix to let this initial permutation ``jump'' to the identity $e$. As each such ``jump'', corresponding to a single transposition, carries a cost $\mc{O}(\chi^{-1})$ (see Eq.~\eqref{eq:transfer}), we find that bringing an initial pairing $\tau$ to the identity $e$ introduces, relative to the Haar case, an overall factor of  
\begin{align}\label{eq:sub}
    \begin{split}
     & \underbrace{\chi^{-k}}_{\boxed{\scriptstyle 1}} \underbrace{\left( d^{2k-2} \right)^m}_{\boxed{\scriptstyle 2}} \underbrace{\chi^{2k} \chi^{-k-2} }_{\boxed{\scriptstyle3}} =\left( \frac{d^{N (1-k^{-1})}}{\chi} \right)^{2k} \, .       
    \end{split}
\end{align}
The term $\boxed{\scriptstyle 1}$ is the cost of the $k$ transpositions, term $\boxed{\scriptstyle 2}$ is the relative gain from placing the identity $e$ at each site, and term $\boxed{\scriptstyle 3}$ arises from the different right boundary condition. We therefore define a scaling limit (s.l.) in which $N, \chi \to \infty$ while keeping fixed the ratio 
\begin{align}\label{eq:xx}
    \begin{split}
     x = d^{N (1-k^{-1})} /\chi \, .       
    \end{split}
\end{align}
In this s.l., the only configurations that give a non-vanishing contribution to $\overline{\mu_k}$ are those in which, when going from left to right in Eq.~\eqref{eq:transfer}, off-diagonal terms of $T$ ``undo'' one after the other distinct pairs of the initial permutation. For example, starting from $\tau=(12)(34)(56)\dots$, one may first transition to $(1)(2)(34)(56)\dots$ and then to $(1)(2)(3)(4)(5)(6)\dots$. Intermediate permutations consist of an even number of singletons together with some remaining pairings. Now, in general, we can expand the transfer matrix by using Eq.~\eqref{eq:Wgexp} as
\begin{align}
\begin{split}
    T_{\sigma \pi} &= \sum_{\rho \in S_{2k}} d^{\#(\sigma)} \frac{M(\sigma^{-1}\rho)}{(d \chi)^{4k-\#(\sigma^{-1}\rho)}} d^{\#(\rho) + 2 \cdot \mathbf{1}_E(\rho) - 2}   \\ 
    & \qquad \quad \, \,  \times \chi^{2k-\#(\rho^{-1}\pi)}\left(1 + \mc{O}(\chi^{-1}) \right)  .
\end{split}
\end{align}
If both $\sigma$ and $\pi$ have the above form (singletons and pairings) and $\pi$ is obtained from $\sigma$ by undoing $\beta$ pairings, then the sum over $\rho$ is dominated by permutations that undo only a subset of size $0 \leq b \leq \beta$ of these pairings. Using this observation, we find
\begin{align}\label{eq:transferUnitaryRMPU} 
    T_{\sigma \pi} &= \frac{d^{2\#(\sigma)-2}}{d^{2k} } \chi^{-\beta} \Bigg[\sum_{b =0}^\beta \binom{\beta}{b}  (-1)^{b}   d^{2 \delta_{b,0} 1_{E}(\sigma)}       \Bigg] \left(1 + \mc{O}(\chi^{-1}) \right) ,
\end{align}
where the term $(-1)^b$ comes from evaluating the function $M(\cdot)$ in Eq.~\eqref{eq:Wgexp} for the intermediate permutations $\rho$. For $\beta = 0$, one obtains $T_{\tau\tau} = 1 + \mc{O}(\chi^{-1})$ when $\sigma = \pi = \tau$ (with $\tau$ a perfect pairing), and $T_{ee} = d^{2k-2}\bigl(1 + \mc{O}(\chi^{-1})\bigr)$ when $\sigma = \pi = e$. For $\beta > 0$, the bracketed sum vanishes unless $\sigma$ is a full pairing, such that $\sigma = \tau$ and $\pi = e$ instead yields: $T_{\tau e} = \frac{d^{2k-2}}{d^{2k} } \chi^{-k}  \sum_{b=0}^{\beta} \binom{\beta}{b} (-1)^{b} d^{2\delta_{b,0}} = d^{-2} (d^{2} - 1)$. Consequently, in our scaling limit, the only contributing terms are those corresponding to a full $k$-fold jump from $\tau$ to $e$ at a single site. Summing over the possible positions of this jump produces the following geometric series
\begin{equation}\label{eq:geometric_series}
\sum_{\ell=0}^{m-1}d^{\ell(2k-2)} = \frac{d^{m(2k-2)}d^2-d^2}{d^{2k}-d^2}\simeq \frac{d^{m(2k-2)+2}}{d^{2k}-d^2} \,.
\end{equation}
Combining all these factor with arrive at 
\begin{equation}\label{eq:final_formula_unitary}
    \overline{\mu_k}
    \stackrel{\rm s.l.}{=}
     \overline{\mu_k}^{\rm Haar}
    \left[
    1+
    C_k  \, x^{2k}
    \right],
\end{equation}
with $C_k=({d^2-1})/({d^{2k}-d^2})$, as reported in Eq.~\eqref{eq:unitaryRMPU}.

\subsection{RMPU with Depolarizing Noise}
We consider now the RMPU ensemble with finite noise $\gamma>0$, following the architecture shown in Fig.~\ref{fig:architectures} (right). We begin by noticing that in the case of a Haar-unitary gate $U$ followed by the depolarizing channel $\mathcal{N}_{\gamma}(O) = (1-\gamma) O + \Id  \cdot \Tr[O]/ \Tr[\Id]$, an analogue of Eq.~\eqref{eq:weingarten} still holds:
\begin{align}\label{eq:weingarten_noisy}
\begin{split}
\mathbb{E}_{U \sim \Haar(q)} \left[ \left( \mathcal{N}_{\gamma} \!\cdot  \! (U^* \! \otimes U) \right)^{\otimes n} \right] =\! \sum_{\pi, \sigma \in S_n} \! \tilde{\Wg}_{\pi, \sigma}(q,\gamma)\, |\pi\rrangle \llangle \sigma|  ,
\end{split}
\end{align}
where the modified Weingarten coefficients $\tilde{\Wg}_{\pi, \sigma}(q,\gamma)$ include the effect of noise and reduce to the conventional Weingarten for $\gamma=0$.
Their explicit form is~\cite{sauliere2025noisyq}:
\begin{align}\label{eq:noisyWg}
\begin{split}
\tilde{\mathrm{Wg}}_{{\pi,\sigma}}=
\sum_{i=0}^{\mathrm{n_F}(\pi,\sigma)}
&\binom{\mathrm{n_F}(\pi,\sigma)}{i} 
\frac{\gamma^i}{q^i}
(1-\gamma)^{n-i}
\mathrm{Wg}_{\tilde{\pi}(i),\,\tilde{\sigma}(i)}^{(n-i)}\,. 
\end{split}
\end{align}
where $\mathrm{n_F}(\pi,\sigma)$ denotes the number of common fixed points between $\pi$ and $\sigma$, and $\tilde{\pi}(i)$ and $\tilde{\sigma}(i)$ denote the permutations obtained from $\pi$ and $\sigma$, respectively, by removing $i$ of these common fixed points (exactly which ones is irrelevant, because $\Wg$ depends only on the cycle structure of its argument). Using Eq.~\eqref{eq:weingarten}, these modified Weingarten coefficients can be expanded for large dimensions $q$ as 
\begin{equation}
\tilde{\mathrm{Wg}}_{\pi,\sigma}(q) = \frac{M(\pi^{-1}\sigma)}{q^{2n-\#(\pi^{-1}\sigma)}} (1-\gamma)^{n-n_F(\pi,\sigma)} \left(1 + \mc{O}(q^{-2}) \right) \, ,
\end{equation}
from which we can see that the noise $\gamma>0$  introduces a bias that favors permutations with the largest number of fixed points, i.e.\ it biases the system toward the identity permutation $e$. 

The similarity between this expansion and the noiseless analogue (Eq.~\eqref{eq:Wgexp}) means that the derivation presented in the previous section can be reused with only minor changes. Specifically, in the s.l., the only jumps that survive are again full $k$-jumps from $\tau$ to $e$, but this time we must include an extra factor $(1-\gamma)^{-2k}$ for each site where the pairing $\tau$ is present. Eq.~\eqref{eq:geometric_series} becomes therefore
\begin{align}
\sum_{\ell=0}^{m-1} (d^{2k-2} (1-\gamma)^{-2k})^{\ell} \stackrel{\rm s.l.}{=}\frac{d^{m(2k-2)} d^2 (1-\gamma)^{-2km}}{d^{2 k}(1-\gamma)^{-2k}-d^2} \, . \label{eq:geoSeries}
\end{align}
We therefore obtain
\begin{equation}\label{eq:formula_noise_neilnew1}
    \overline{\nu_k} \stackrel{\rm s.l.}{=}
     \overline{\mu_k}^{\rm Haar} (1-\gamma)^{2km} \left[1 +    C_k(\gamma) \left( \frac{x}{(1-\gamma)^{m}}\right)^{2k} \right] \, ,
\end{equation}
with $C_k(\gamma) := \big(d^2-1\big) / \big(d^{2k} (1-\gamma)^{-2k} - d^2\big)$, as presented in Eq.~\eqref{eq:formula_noise_neilnew}. As a consistency check, we indeed find that for $\gamma=0$, Eq.~\eqref{eq:formula_noise_neilnew1} reduces to Eq.~\eqref{eq:final_formula_unitary}.

\section{Statistical model for the brickwork circuit} \label{ap:stat_mech}
We describe \new{in more detail} the statistical mechanical model arising from averaging $2k$ copies of the noisy random brickwork circuit, \new{proving the construction summarized in Section~\ref{sec:stat_mech}.} 

Consider the elementary building block of the circuit, consisting of three noisy gates acting on four contiguous qudits $i-1, i, i+1, i+2$. The object we need to compute is the average over the Haar-random gates of this block, replicated $2k$ times, i.e.
\begin{equation}
    \mathbb{E}_{U_{i-1,i},U_{i,i+1},U_{i+1,i+2} \sim \Haar(d^2)} \Bigg[ \Big(\includegraphics[scale=0.55, valign=c]{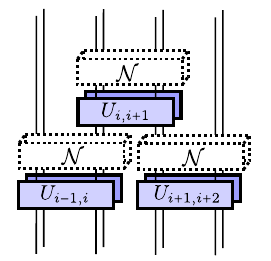} \Big)^{\otimes 2k} \Bigg]
\end{equation}
We can then apply Eq.~\eqref{eq:weingarten_noisy} on each noisy gate. For example, for the gate acting on $i,i+1$, we obtain
\begin{align}
\begin{split}
\mathbb{E}_{U_{i,i+1} \sim \Haar(d^2)}&\left[ \left( \mathcal{N}_{\gamma} \cdot (U_{i,i+1} \otimes U_{i,i+1}^*) \right)^{\otimes 2k} \right]  \\ = \sum_{\rho, \delta \in S_n} & \tilde{\Wg}_{\rho, \delta}(d^2,\gamma)\, |\rho\rrangle_{i} |\rho\rrangle_{i+1} \llangle \delta|_{i} \, \llangle \delta|_{i+1}  \, ,
\end{split}
\end{align}
where we explicitly indicate the qudit associated with each permutation state.  Eq.~\eqref{eq:weingarten_noisy} applied on the gates $U_{i-1,i}$ and $U_{i+1,i+2}$ yields other permutation states $|\sigma \rrangle_{i}$ and $|\pi \rrangle_{i+1}$, which then are contracted with $\llangle \delta|_{i}$ and $ \llangle \delta|_{i+1}$ respectively; this gives: $G_{\delta \sigma}(d) = \llangle \delta | \sigma \rrangle$ and $G_{\delta \pi}(d) = \llangle \delta | \pi \rrangle$. Putting this all together, this defines the following weights associated with an elementary plaquette of three permutations:
\begin{equation}
J_{\sigma \pi \rho}(\gamma) := \sum_{\delta \in S_n} \tilde{\Wg}_{\rho \delta}(d^2;\gamma) G_{\delta \sigma}(d) G_{\delta \pi}(d) =
\begin{tikzpicture}[scale=1.2, baseline=15pt]
\coordinate (A) at (-0.5,0);
\coordinate (B) at (0.5,0);
\coordinate (C) at (0,+0.866); 
\draw[thick, fill=green!20] (A) -- (B) -- (C) -- cycle;
\filldraw[fill=white, draw=black] (A) circle (1.5pt);
\filldraw[fill=white, draw=black] (B) circle (1.5pt);
\filldraw[fill=white, draw=black] (C) circle (1.5pt);
\node[scale=1.] at (-0.5,-0.2) {$\sigma$};
\node[scale=1.] at (0.5,-0.2) {$\pi$};
\node[scale=1.] at (0,+1.066) {$\rho$};
\end{tikzpicture} \, .
\end{equation}
The full calculation of $\overline{\mu_k}$ can then be represented as the following lattice 
\begin{equation}\label{eq:muk_lattice}
\raisebox{47pt}{$\overline{\mu_k} = \qquad $}  \includegraphics[width=0.655\linewidth]{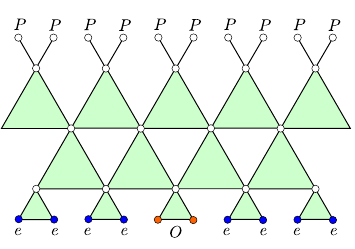}
\end{equation}
where a permutation is associated to each vertex $\begin{tikzpicture}[scale=1.2, baseline=-3pt]\filldraw[fill=white, draw=black] (0,0) circle (1.5pt); \end{tikzpicture}$, and all permutations are implicitly summed over. Moreover, we introduce the notation: $\begin{tikzpicture}[scale=0.7, baseline=-12pt]
\coordinate (A) at (-0.5,0);
\coordinate (B) at (0.5,0);
\coordinate (C) at (0,-0.866); 
\draw[thick] (A) -- (C) -- (B);
\filldraw[fill=white, draw=black] (A) circle (1.5pt);
\filldraw[fill=white, draw=black] (B) circle (1.5pt);
\filldraw[fill=white, draw=black] (C) circle (1.5pt);
\node[scale=1.] at (-0.5,+0.2) {$\sigma$};
\node[scale=1.] at (0.5,+0.2) {$\pi$};
\node[scale=1.] at (0,-1.1) {$\rho$};
\end{tikzpicture} = \delta_{\sigma \rho} \delta_{\pi \rho}$, and $\begin{tikzpicture}[scale=1.2, baseline=-3pt]\filldraw[fill=blue, draw=black] (0,0) circle (1.5pt); \end{tikzpicture}$ to represent the identity permutation $e$. At the top and bottom boundaries, the permutation states are weighted by contraction with the Pauli operators $P$ and with the initial operator $O$ respectively, yielding: $\begin{tikzpicture}[scale=1.2, baseline=0pt]
\coordinate (A) at (0,0);
\node[scale=0.9] at (0,+0.17) {$P$};
\node[scale=0.9] at (0,-0.16) {$\sigma$};
\filldraw[fill=white, draw=black] (A) circle (1.5pt);
\end{tikzpicture}
= \sum_P \llangle P^{\otimes 2k} | \sigma \rrangle = d^{\#(\sigma) + 2 1_E(\sigma)}$ and $\begin{tikzpicture}[scale=1.2, baseline=0pt]
\coordinate (A) at (0,0);
\node[scale=0.9] at (0,+0.15) {$\sigma$};
\node[scale=0.9] at (0,-0.18) {$O$};
\filldraw[fill=orange, draw=black] (A) circle (1.5pt);
\end{tikzpicture}
= \llangle O^{\otimes 2k} | \sigma \rrangle = d^{\#(\sigma)} 1_E(\sigma)$. Eq.~\eqref{eq:muk_lattice} can be understood as the partition function of an Ising-like model on a triangular lattice, with permutations $\sigma \in S_n$ taking the role of Ising variables (strictly speaking, this interpretation is possible only when the weights $J_{\sigma \pi \rho}$ are positive definite, which occurs for large enough $d$~\cite{Bao_2020}). Crucially, the weights are maximized by the uniform configuration $\sigma=\pi=\rho$. In this case they are $J_{\sigma \sigma \sigma} = (1-\gamma)^{2k - n_F(\sigma)}$, with $n_F(\sigma)$ the number of fixed points of the permutation $\sigma$. 
These facts show that: i) the statistical model is ferromagnetic, with domain walls paying an energetic cost of $\mathcal{O}(d^{-1})$ per plaquette they pass through; ii) the noise plays the role of an external magnetic field polarizing over the permutations with the largest number of fixed points, i.e.\ over the identity $e$.

Another important property of the weights is that $J_{e e \rho}(\gamma) = [G(d^2) \tilde{\Wg}(d^2;\gamma)]_{e \rho} = \delta_{e \rho}$. This shows that when the two permutation states on the bottom of the plaquette are both the identity $e$, the third one is forced to be the identity as well. This fact imposes a lightcone structure:
\begin{equation}\label{eq:lightcone}
\raisebox{47pt}{$\overline{\mu_k} = \qquad $}  \includegraphics[width=0.655\linewidth]{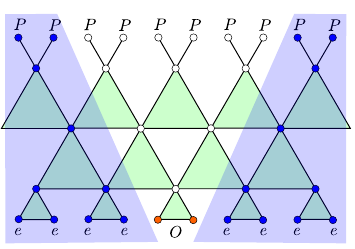}
\end{equation}
where permutations outside the cone are fixed to be the identity. This light cone arises from the finite propagation speed of the initially localized operator $O$, a direct consequence of the circuit structure. In the main text, we discuss the relevant permutation configurations that emerge \emph{within this cone}; in particular, we argue that in the large-noise regime, the dominant contributions come from configurations that further shrink the cone (see Fig.~\ref{fig:cartoon}, right).

\section{Numerical methods}\label{ap:numerics}
To perform numerics for the brickwork circuit model, we used both the \emph{Replica Tensor Network} (RTN) approach and exact state vector simulation of the time-evolved operator. RTN is a numerical technique to evaluate the partition function of the statistical-mechanics model arising from replicating and averaging a random circuit. In fact, Eq.~\eqref{eq:muk_lattice} can be viewed as a two dimensional tensor network, where the degrees of freedom are $(2k)!$-dimensional (i.e.\ they correspond to permutations). This tensor network can then be contracted using standard techniques. Specifically, one can contract it from bottom to top, using Matrix Product States (MPS) to represent the `state' of the system.
In the noiseless case ($\gamma=0$), we used RTN to compute the second moment $\overline{\mu_2}$ in Fig.~\ref{fig:plots1} (the MPS used had bond dimensions up to $\chi_{\mathrm{MPS}}=1024$). The main limitations of RTN are: i) only $k=2$ is easily doable, since for $k=3$ the degrees of freedom already have dimension $6! = 720$; ii) in the noisy case ($\gamma > 0$), one cannot compute the normalized moments $\overline{\mu_k}$ because the normalization introduces a non-polynomial dependence on the gates $U$; only the unnormalized moments $\overline{\nu_k}$ are computable. For this reason, we also employ exact state vector simulation of the time-evolved operator. That is, we use the vectorized representation of the operator $|O\rrangle$ as a vector of length $d^{2N}$ and evolve it by applying the circuit gate by gate. For a single gate, the update takes the form: $|O\rrangle \to \mathcal{N} \cdot (U \otimes U^*) |O\rrangle$. At the end, we obtain the final operator $|O_U\rrangle$ and we rotate it to the Pauli basis to obtain the overlaps $\llangle P|O_U\rrangle = \Tr[P O_U]$, from which we directly compute the moments. This method allows us to compute both the normalized moments $\mu_k$ and the unnormalized moments $\nu_k$. Circuit average can be achieved by collecting data for sufficiently many random circuit realizations. We apply this method to obtain the third moment $\overline{\mu_3}$ (shown in Fig.~\ref{fig:plots1}) and the noisy results for $\overline{\nu_2}$ (presented in Fig.~\ref{fig:plots2}). In both cases, we collect $1000$ circuit realizations.

\new{
\section{Haar average of the relative error}\label{sec:haaraveragerelativeerror}
Here we justify the form of the square relative error in expectation values analyzed in Section.~\ref{sec:simtransition}. 
The squared relative error on a fixed reference state $|\psi\rangle$ is defined as
\begin{equation}
    \langle\psi|\Delta O_U|\psi\rangle^2/\langle\psi|O_U|\psi\rangle^2,
\end{equation}
with $\Delta O_U := O_U-\tilde O_U$. We now consider Haar-random states $|\psi\rangle$. Such a state can be written as $|\psi\rangle=z/\|z\|$, with $z\in\mathbb C^D$ a vector of i.i.d.\ standard complex Gaussians. The norm of this vector is $\|z\|^2=D\,[1+\mathcal O(D^{-1/2})]$. Let $A$ now be any traceless Hermitian operator, written as $A=V\Lambda V^\dagger$, with $\Lambda$ the diagonal matrix of its eigenvalues $\{\lambda_i\}$. Then $\langle\psi|A|\psi\rangle\simeq D^{-1}\sum_i\lambda_i|w_i|^2$,
where $w=V^\dagger z$ is again i.i.d.\ Gaussian by unitary invariance. This is a sum of $D$ independent
terms, with means $\lambda_i/D$ (summing to zero since $\Tr[A]=0$). By the central limit theorem (Lyapunov
formulation), $\langle\psi|A|\psi\rangle$ is therefore a real Gaussian variable with zero mean and variance $\Tr[A^2]/D^2$, provided no single eigenvalue dominates, i.e.\ $\Tr[A^4]\ll(\Tr[A^2])^2$.}

\new{This condition holds for both $A=O_U$, and $A=\Delta O_U$, and their linear combinations. Indeed writing
$A=\sum_{P\neq \Id}c_P\,P$, we have $\Tr[A^2]=D\sum_P c_P^2$. When only $n\ll D$ Pauli strings dominate,
$\Tr[A^4]=\sum_i\lambda_i^4\le\|A\|_\infty^2\Tr[A^2]$, with $\|A\|_\infty^2=\max_i\lambda_i^2$ and $\|A\|_\infty  \le \sum_P|c_P| \, \|P\|_\infty=\sum_P|c_P|$ give
\begin{equation}
  \frac{\Tr[A^4]}{(\Tr[A^2])^2}\le\frac{(\sum_P|c_P|)^2}{D\sum_P c_P^2}\le\frac{n}{D}.
\end{equation}
In the Porter--Thomas regime, instead, the weight is spread over all strings, $c_P^2=\mathcal O(D^{-2})$,
and $A$ behaves as a random matrix, with $\Tr[A^4]/(\Tr[A^2])^2\simeq2/D$. In both cases the ratio is
$\mathcal O(1/D)$, i.e.\ exponentially small in $N$. \\
Hence $a=\langle\psi|O_U|\psi\rangle$ and $b=\langle\psi|\Delta O_U|\psi\rangle$ are jointly Gaussian,
with zero mean, variances 
\begin{equation}
\Ex_{\psi\sim \Haar}[a^2] = \frac{\Tr[O_U^2]}{D^2} \, \qquad  \Ex_{\psi\sim \Haar}[b^2] = \frac{\Tr[\Delta O_U^2]}{D^2}
\end{equation}
and covariance $\Tr[O_U\Delta O_U]/D^2$. Since they are correlated, we split $b$ into a part proportional to $a$ and a
remainder,
\begin{equation}
  b=\frac{\Tr[O_U\Delta O_U]}{\Tr[O_U^2]}\,a+u ,
\end{equation}
where the coefficient is chosen so that $u$ is uncorrelated with $a$. For jointly Gaussian variables, uncorrelated means independent, so $u$ is a zero-mean Gaussian independent of $a$. Now we have
\begin{equation}
  \frac{\langle\psi|\Delta O_U|\psi\rangle}{\langle\psi|O_U|\psi\rangle}=\frac{b}{a}
  =\frac{\Tr[O_U\Delta O_U]}{\Tr[O_U^2]}+\frac{u}{a}.
\end{equation}
The ratio of two independent zero-mean Gaussians follows a Cauchy distribution, with scale given by
the ratio of their standard deviations, so that
\begin{align}
\begin{split}
  &\frac{b}{a}\sim\mathrm{Cauchy}\left(\Tr[O_U\Delta O_U]/\Tr[O_U^2], \, s \right),\\
  &s=\frac{\sqrt{\Tr[O_U^2]\,\Tr[\Delta O_U^2]-\Tr[O_U\Delta O_U]^2}}{\Tr[O_U^2]} \, .    
\end{split}
\end{align}
Hence $b^2/a^2$ has infinite mean, and the average (square) relative error 
\begin{equation}
  \Ex_{\psi\sim\Haar}\bigg[\frac{\langle\psi|\Delta O_U|\psi\rangle^2}{\langle\psi|O_U|\psi\rangle^2}\bigg]
  =\Ex\bigg[\frac{b^2}{a^2}\bigg]
\end{equation}
diverges. Nevertheless, the quantity 
\begin{equation}
  \varepsilon_U=\frac{\Ex_\psi[b^2]}{\Ex_\psi[a^2]},
\end{equation}
which we compute in
Section~\ref{sec:simtransition} (see Eq.~\eqref{eq:ratio_new}), is finite and gives the typical value of
the relative error for Haar-random states, i.e.\ it sets the median.}

\bigskip

\section*{Acknowledgments}
\new{No AI tools were used for any part of the ideas, analytical results, or writing the paper.
Claude Opus 5.5 (Anthropic) has been used to optimize the code for the cluster and document it.
}
We acknowledge use of the cluster of Coll\`ege de France. 
We thank A. De Luca and A. Sauliere for inspiring discussions and collaboration on related subjects. J.D.N. and G.L. are funded by the ERC Starting Grant 101042293 (HEPIQ) and the ANR-22-CPJ1-0021-01. N.D. and X.T. acknowledge support from DFG under Germany's Excellence Strategy – Cluster of Excellence Matter and Light for Quantum Computing (ML4Q) EXC 2004/2 – 390534769. X.T. further acknowledges support from DFG Collaborative Research Center (CRC) 183 Project No. 277101999 - project B01, and DFG Emmy Noether Programme proposal ``Digital Quantum Matter Out-of-Equilibrium'' No. 560726973. 

\section*{Code and Data Availability}
The code and data supporting the findings will be available in Zenodo at publication. 

\section*{Author contributions}
N.D. and G.L designed the project, with assistance from J.D.N. and X.T.. N.D. and G.L. performed the analytical calculations for the RMPU model. G.L. and X.T. conducted all the numerical simulations. J.D.N. and G.L. conceived the membrane picture in the brickwork circuit. N.D. derived the simulability bounds. All authors contributed equally to writing the paper.

\bibliography{references}

\end{document}